\newcommand{\psup}{\ensuremath{P_{\rm{sup}}}}
\newcommand{\pspin}{\ensuremath{P_{\rm{spin}}}}
\newcommand{\pdot}{\ensuremath{\dot{P}_{\rm{spin}}}}
\newcommand{\dpspin}{\ensuremath{\langle \dot{P}_{\rm{spin}} \rangle}}

\newcommand{\rxte}{{\it RXTE}}
\newcommand{\rxteasm}{{\it RXTE/ASM}}
\newcommand{\rxtepca}{{\it RXTE/PCA}}
\newcommand{\chandra}{{\it Chandra}}
\newcommand{\INTEGRAL}{{\it INTEGRAL}}
\newcommand{\rosat}{{\it ROSAT}}
\newcommand{\bepposax}{{\it BeppoSAX}}
\newcommand{\XMM}{{\it XMM}}

\documentclass[fleqn,usenatbib]{mnras}

\usepackage{newtxtext,newtxmath}

\usepackage[T1]{fontenc}
\usepackage{ae,aecompl}

\usepackage{graphicx}	
\usepackage{amsmath}	
\usepackage{amssymb}	

\defcitealias{clarkson2003a}{Paper~I}
\defcitealias{clarkson2003b}{Paper~II}
\defcitealias{od01}{OD01}
\defcitealias{inam2010}{I10}
\defcitealias{neilsen04}{N04}

\title[X-ray spin-superorbital correlation in SMC X-1?]{Longterm properties of accretion discs in X-ray Binaries - III. A search for spin-superorbital correlation in SMC X-1}

\author[Kristen Dage, Will Clarkson, Phil Charles, Silas Laycock \& I-Chun Shih]
{Kristen C. Dage,$^{1, 2}$\thanks{E-mail: kcdage@msu.edu}
William I. Clarkson,$^{1}$
Philip A. Charles$^{3,4, 5}$,
\newauthor
Silas G. T. Laycock$^6$~and I-Chun Shih$^7$
\\
$^{1}$Department of Natural Sciences, University of Michigan-Dearborn, 4901 Evergreen Rd. Dearborn, MI, 48128\\
$^{2}$Department of Physics and Astronomy, Michigan State University, East Lansing, MI, 48824, USA\\
$^3$Astrophysics, Department of Physics, University of Oxford, Clarendon Laboratory, Parks Road, Oxford, OX1 3PU, UK\\
$^4$Physics \& Astronomy, University of Southampton, SO17 1BJ, UK\\
$^5$Leverhulme Emeritus Fellow\\
$^6$Department of Physics \& Applied Physics, University of Massachusetts, Lowell, MA 01854, USA\\
$^7$GEPI, Observatoire de Paris, Universit\'{e} PSL, CNRS, 5 Place Jules Janssen, 92190 Meudon, France
}

\date{Accepted XXX. Received YYY; in original form ZZZ}

\pubyear{2018}

\begin{document}
\label{firstpage}
\pagerange{\pageref{firstpage}--\pageref{lastpage}}
\maketitle

\begin{abstract}

Thanks to long-term X-ray monitoring, a number of interacting binaries are now known to show X-ray periodicities on timescales of tens to hundreds of binary orbits. In some systems, precession of a warped accretion disc is the leading model to explain the superorbital modulation. The High Mass X-ray Binary SMC X-1 showed two excursions in superorbital period (from $\sim$60~d to $\sim$45~d) during the 1996-2011 interval, suggesting that some characteristic of the accretion disc is varying on a timescale of years. Because its behaviour as an X-ray pulsar has also been intensively monitored, SMC X-1 offers the rare chance to track changes in both the accretion disk and pulsar behaviours over the same interval. We have used archival X-ray observations of SMC X-1 to investigate whether the evolution of its superorbital variation and pulse period are correlated. We use the 16-year dataset afforded by the {\it RXTE}~All-Sky Monitor to trace the behaviour of the warped accretion disc in this system, and use published  pulse-period histories to trace the behaviour of the pulsar. While we cannot claim a strong detection of correlation, the first superorbital period excursion near MJD 50,800 does coincide with structure in SMC X-1's pulse period history. Our preferred interpretation is that the superorbital period excursion coincides with a change in the long-term spin-up rate of the SMC X-1 pulsar. In this scenario, the pulsar and the accretion disc are both responding to a change in the accretion flow, which the disc itself may regulate. 

\end{abstract}




\section{Introduction}\label{sec:intro}

With orbital periods typically a few days or shorter \citep[e.g.][]{liu06},  X-ray binaries allow behaviours to be charted over hundreds of orbital timescales through multi-year X-ray monitoring, offering a window into the accretion/outflow process which is important to astrophysical objects on all lengthscales \citep[e.g.][]{fkr02, charlesCoe06}. As suggested by X-ray monitoring datasets on timescales of years-decades, a number of X-ray binaries are now suspected to undergo accretion disc warping, in which the accretion disc warps out of the plane and may precess, giving rise to observed high/low cycles in X-ray brightness on a timescale of tens of days \citep[e.g.][]{charlesKotze10}. 

The leading candidate mechanism to produce the warp is {\it radiation-driven warping} \citep{pringle1996}, in which the interception and re-emission by the accretion disc of radiation from the accretor leads to a perturbative force out of the plane. For the more luminous accretors at the centre of sufficiently large accretion discs, a warped configuration becomes stable \citep{wijers1999} and can precess. Under this model, the configuration of the warp, and thus the superorbital modulation, should depend on the binary separation (a proxy for the disc radius), with more widely separated binaries producing more complicated warp configurations, or possibly a competition of modes (\citealt{od01}, hereafter \citetalias{od01}). Comparison of superorbital X-ray modulations of a sample of systems chosen to trace the warp modes of \citetalias{od01} suggested that the X-ray binaries as a class might indeed follow this scheme (\citealt{clarkson2003a}, \citealt{clarkson2003b}, hereafter \citetalias{clarkson2003a} \& \citetalias{clarkson2003b}, respectively). Smoothed-particle hydrodynamics simulations of accretion discs in a variety of realistic configurations added further support to the radiation-driven warping model \citep{foulkes2006, foulkes2010}; when the central radiation-source is switched on in the simulation, the accretion disc does indeed warp and precess.

Because it is the interception of X-ray luminosity that drives the warp in the first place, the configuration of the warp should be sensitive to the mass transfer rate onto the central accretor \citep{od01, foulkes2010}. An X-ray binary in which X-ray superorbital modulation is observed to vary, {\it and} in which variations in $\dot{M}$ can be observed, would therefore provide a direct test of the disc/accretor link for warped discs.

The well-studied X-ray binary SMC X-1 is the perfect system in which to probe this connection. A superorbital X-ray modulation at $\sim$60 days has long been observed (\citealt{bonnet-bidaud1981} provides an early discussion of superorbital modulation of this source). X-ray spectroscopic evidence suggests this variation is likely due to repeated obscuration, just as would be expected from a precessing disc-warp (see, e.g. \citealt{woj98, vrtilek05}, as well as \citealt{blondinWoo1995}). Optical monitoring supports this general model \citep{coe2013}. 

The superorbital modulation from SMC X-1 was suspected to vary at least as early as 1998 thanks to the first year of X-ray monitoring by \rxte~\citep{woj98}. Further monitoring has led to improved characterization of this variation (e.g. \citealt{ribo01}; \citetalias{clarkson2003a}; \citealt{trowbridge2007}; \citealt{kotzeCharles2012}; \citealt{hu2013}). Whether the detected period of modulation varies smoothly or discretely, on at least two occasions the superorbital modulation has been recorded to be as short as 45 days. We identify in particular two superorbital period excursions: the first superorbital excursion takes place between approximately MJD 50,500-51,500, with period minimum taking place near MJD 50,800. The second, slightly shorter superorbital excursion takes place between approximately MJD 53,500-54,500, with period minimum taking place near MJD 54,000 \citep[e.g.][]{kotzeCharles2012, trowbridge2007}. SMC X-1 lies in the region of (binary radius, mass ratio) space in which a competition of modes might be expected \citepalias{od01}. It is the best current candidate for a system in which an accretion disc warp shows strong variation in its configuration.

Some indication of a link between superorbital modulation and $\dot{M}$~has been claimed from long-term X-ray monitoring alone. For example, SMC X-1 shows a weak correlation between the superorbital period and superorbital modulation amplitude (in the sense that longer modulation period would suggest greater modulation amplitude), although the link between rms amplitude and X-ray flux (tracing $\dot{M}$) is unclear \citep{hu2011}. {\it If} this amplitude indeed relates to $\dot{M}$~directly, then this claimed correlation would be a point of similarity with Her X-1, the prototypical X-ray binary exhibiting stable accretion disk warping \citep[varying in the range 33-37 days, e.g.][]{jurua2011}, for which a correlation has also been measured between its superorbital modulation period and the average X-ray flux over the superorbital cycle \citep{leahy2010a}.

SMC X-1\footnote{SMC X-1 is object SXP 0.72 in the \citet{coe05} catalog of X-ray pulsars in the SMC, as well as in subsequent catalogs using the same nomenclature (e.g. \citealt{Yang2017}).} is also a very well-measured X-ray pulsar, with the evolution of the X-ray pulse profile offering further support for a warped accretion disc scenario. Its X-ray pulse profile is energy-dependent; the ``hard" component (at energies $> 2.0$~keV) is double peaked and does not show strong variation in profile shape with superorbital phase (e.g. \citealt{neilsen04} and references therein). However, a ``soft" (energies $< 1.0$~keV) component has also been detected, whose profile does vary with superorbital phase in a manner consistent with reprocessing of the hard component by the inner regions of a warped accretion disc \citep{hickox2005}.  

The pulse period history of SMC X-1 suggests its neutron star accretor has consistently been spun up over all five decades in which X-ray pulse-period measurements have been taken \citep{naikPaul2004}, which suggests a persistent accretion disc despite its HMXB nature (the companion is a 17M$_{\odot}$~B0 supergiant; \citealt{vanderMeer2007}).\footnote{SMC X-1 is not the only HMXB pulsar to show a near-monotonic spin-up trend over many years; at least a dozen additional systems show mostly negative period derivative within the SMC alone \citep[e.g.][]{klus14, Yang2017}.} 

SMC X-1 therefore offers a crucial test-case in which changes in {\it both} the superorbital period and the accretion torque on the pulsar can be measured. We have therefore undertaken a side-by-side comparison of changes in  superorbital period \psup~(from publicly-available RXTE/ASM data) and  pulse period \pspin~(from published pulse periods). Our aim is to learn whether changes in the putative accretion disc-warping and precession are indeed accompanied by changes in the accretion-induced torque on the neutron star.

This paper is the third in a series examining long-term X-ray periodicities in X-ray binaries and comparing them to the radiation-driven warping framework. \citetalias{clarkson2003a} demonstrated the variation in \psup~from  SMC X-1, building on indications from \citet{woj98} and \citet{ribo01}. In \citetalias{clarkson2003b} the analysis was extended to include Her X-1, LMC X-4 and Cyg X-2, covering a range of locations in the (binary radius - mass ratio) diagram of \citetalias{od01} and suggesting that the radiation-driving framework of \citetalias{od01} is consistent with measurements (\citealt{boyd04} provide a different but complementary interpretation). Both papers used \rxte~All Sky Monitor data over the interval 1996 December - 2002 August.  

This paper is organized as follows: the datasets used are introduced in Section \ref{sec:obsdata}, and their characteristics relevant to this study summarized; Section \ref{sec:methods} communicates the techniques used to isolate changes in the superorbital and spin periods, respectively,  with Appendices \ref{app:sec:uncty} \& \ref{ap:timing} presenting detailed methodology on the determination of superorbital period and particularly its measurement uncertainty. The side-by-side comparison of the superorbital- and spin-periods are then presented in Section \ref{sec:results}, with implications for future work discussed in Section \ref{sec:discussion}.

\section{Observational Data}\label{sec:obsdata}

Most of the data used for this study were taken with the Rossi X-ray Timing Explorer (\rxte) satellite, using All-Sky Monitor data (Section \ref{ssec:data_asm}) to chart the superorbital modulation and its variation, and the \citet[][hereafter \citetalias{inam2010}]{inam2010} compilation of pointed observations with the Proportional Counter Array (PCA) to trace evolution in the pulse period. However, the pulse periods are not exclusively from RXTE; in the \rxte~era, pulse period measurements from \rxte, \bepposax, \chandra, \INTEGRAL~and \XMM~are considered in the analysis, while ultimately the published pulse period history of SMC X-1 includes measurements from sixteen missions (\citetalias{inam2010}, \citealt{neilsen04} and Section \ref{ssec:data_pca}).

\subsection{Superorbital modulation: \rxteasm}\label{ssec:data_asm}

\rxte~carried an All-Sky Monitor (ASM), which provided regular coverage of the entire sky at $1.3 - 12.1$~keV energies. It consisted of three scanning detectors whose pointing and flux history were resolved into roughly (0-20)$\times$90s measurements of individual sources per day, with timing accurate to within a thousandth of a day \citep{levine1996}. Crude spectral information is also available through three energy channels (1.3-3.0 keV, 3.0-5.0 keV, and 5.0-12.1 keV), although we do not use this channel-by-channel information in the present analysis. In total, \rxteasm~accumulated just under sixteen years' monitoring on SMC X-1 (or about 1,494 binary orbits).

We considered the entire \rxteasm~time baseline for SMC X-1 (MJD 50,088.35 - 55,902.96, or 1996 Jan 06 - 2011 Dec 07) in our analysis, although the coverage degrades somewhat in the final year of operation so that (for example) superorbital minima become difficult to distinguish; see e.g. Figure \ref{fig:lcurve}. Following previous practice (e.g. \citealt{levine1996}, \citetalias{clarkson2003a}, \citealt{trowbridge2007}), we filtered the ASM data for background ($< 10.0$~counts s$^{-1}$) and on the quality of the least-squares fit to the flux (keeping points with reduced chi-square statistic $\chi_\nu^2 < 1.5$). This filtering removed about 7\% of the datapoints (cutting the sample from 60,246 down to 56,263  measurements).
 
\subsection{Pulse periods: \rxtepca~and other missions}\label{ssec:data_pca}

SMC X-1 has been observed many times with pointed X-ray observations ($> 160$~times with \rxtepca~alone; \citetalias{inam2010}), allowing its $\simeq 0.71$~second pulse period to be tracked on a timescale of decades. \citet{woj98} presented a pulse period history of SMC X-1 over the interval 1971 January - 1996 January. Since then, considerably more pointed X-ray observations have taken place (e.g. \citetalias{inam2010}, \citealt{raichurPaul10} and \citealt{falanga15}). A complete pulse-period history of SMC X-1 including all recent observations from all missions does not yet appear to be available. 

Instead, we rely heavily on the \citetalias{inam2010} compendium of pulse period measurements, which considers $\sim$130~\rxtepca~pointings totaling 250ks over the time interval 1996 January-2003 December, as well as a handful of pointed observations taken with \rosat, \bepposax, \chandra~and \INTEGRAL. This provides sufficient coverage to track the pulse period history of SMC X-1 over its  first superorbital excursion near MJD 50,800 and its subsequent behaviour (e.g. \citetalias{clarkson2003a}), and to make predictions for the behaviour of SMC X-1's spin period during the second (at MJD $\approx$ 54,000, e.g. \citealt{trowbridge2007})  and further superorbital period excursions.

Full details of the extraction of the pulse periods themselves can be found in \citetalias{inam2010}. Briefly, the authors first produced an initial estimate for the observed pulse period from the periodogram peak in the barycentre-corrected lightcurve (at energies 3-20~keV, the \citetalias{inam2010} pulse periods correspond to the ``hard" component of the pulse profile; \citealt{hickox2005}). Folding on this initial estimate then produced an empirical pulse profile template, against which 200s-long intervals of the lightcurve were cross-correlated to measure pulse arrival times. Where near-contiguous portions of the data covering several orbits were available, \citetalias{inam2010} deduced orbital elements, allowing an orbit-corrected history of SMC X-1 pulse periods. We extracted the spin frequencies and uncertainties from Table 3 of \citetalias{inam2010}, converting them into pulse periods and uncertainties. 

Seven additional X-ray pulse period measurements tabulated in \citet[][hereafter \citetalias{neilsen04}]{neilsen04} were also included in our analysis; as this second dataset includes five measurements during the two years after the first X-ray superorbital excursion event (three with \chandra, two with \XMM, with two more \rosat~pulse period measurements filling out the coverage before the \rxte~era), this substantially improves X-ray pulse period coverage after the first superorbital period excursion.

The union of the \citetalias{inam2010} and the \citetalias{neilsen04} pulse period measurements completes the set of archival X-ray pulse periods for SMC X-1. Not all archival measurements of otherwise sufficient duration yielded published pulse periods; for example, {\it Chandra} observations 1027-1030 at MJD 51,854, 51,856, 52,024 and 52,025 (all P.I. Vrtilek) were taken during the superorbital low state, while \rxte~observations in 2004 Jan 27-30 inclusive (MJD 53,031-53,035, from program P80078, P.I. Eikenberry) apparently failed to detect the pulsar, despite the source not being at superorbital minimum at the time of these latter observations (Section \ref{ss:dis:moreObs}).

\section{Methods}\label{sec:methods}

Here we describe the methods we used to isolate changes in \psup~and
\pspin. 

\begin{figure*}
\begin{center}
\includegraphics[width=7.0in]{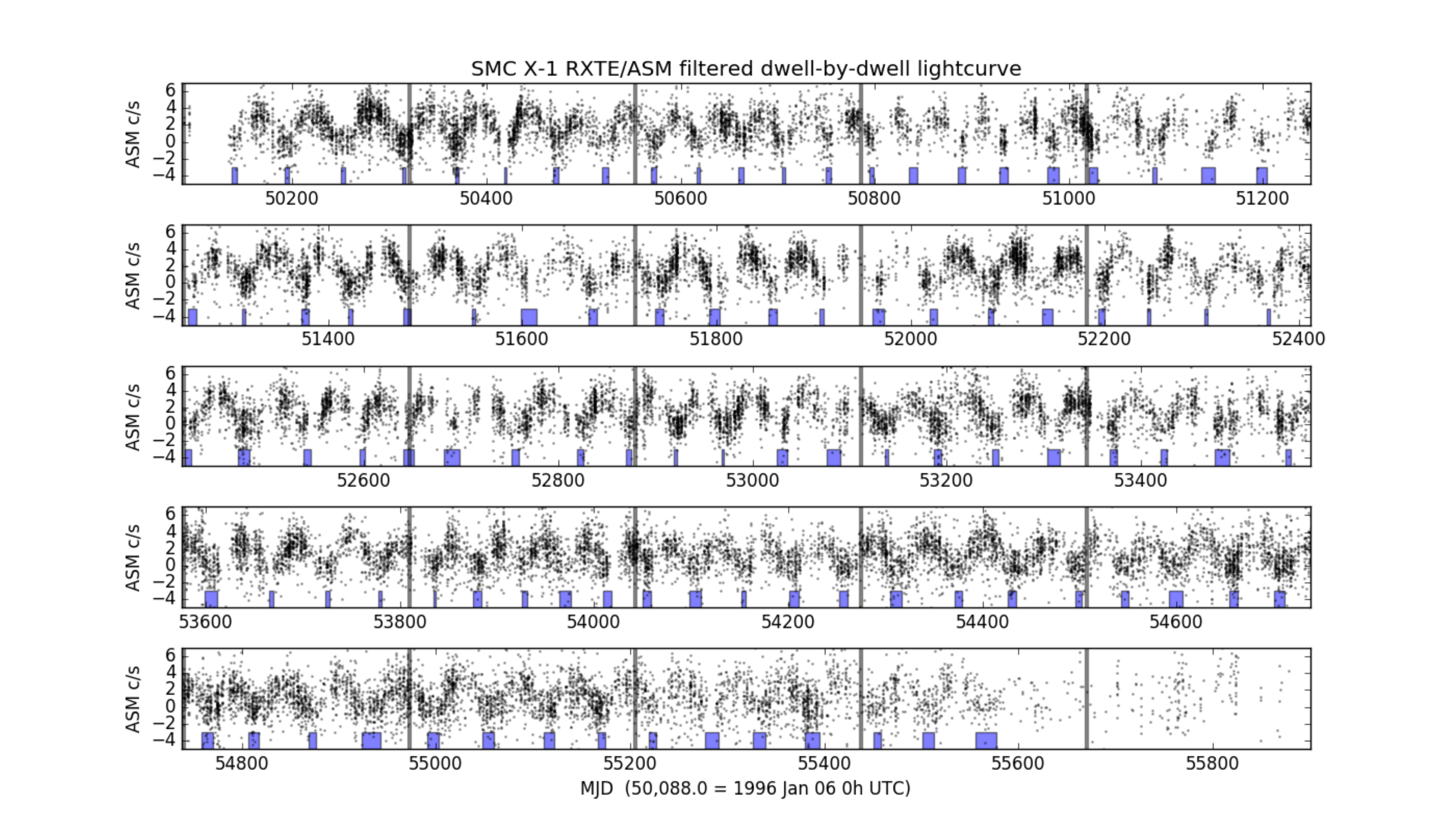}
\caption{The \rxteasm~dwell-by-dwell lightcurve used here. Blue shaded regions show estimated cycle minima estimated using methods similar to those in \citet{trowbridge2007}. This plot shows the interval from the first measurement (at MJD 50,088.0 = 1996 January 06) to MJD 55,901 = 2011 December 05. Vertical gray lines indicate an example division of the lightcurve into 25 equal-length strips.} See Section \ref{ssec:psup}.
\label{fig:lcurve}
\end{center}
\end{figure*}

\subsection{Changes in superorbital variation}\label{ssec:psup}

Figure \ref{fig:lcurve} shows the filtered \rxteasm~lightcurve used here. To track the evolution of superorbital X-ray flux variations from SMC X-1 (whether in \psup, or in the shape of the lightcurve, or both), we employed the Dynamical Power Spectrum (DPS), in which the Lomb-Scargle (LS) periodogram \citep{scargle82} is charted for sub-intervals within the total time interval, allowing the variation in power spectrum behaviour to be charted over time \citep[e.g.][]{kotzeCharles2012}. Other methods that have been used to probe SMC X-1's longterm variability include profile fitting in the time domain to estimate duration of individual variation cycles \citep{woj98,trowbridge2007}, wavelet analysis \citep{ribo01}, and the Hilbert-Huang transform \citep{hu2011, hu2013}; these methods find largely similar behaviour to power spectrum methods, with differences in sensitivity to instantaneous changes in superorbital variation timescale.

The DPS approach suffers from two difficulties that complicate its interpretation, both related to the breaking of the total time baseline into intervals over which to compute each LS power spectrum. Firstly, as pointed out by \citet{trowbridge2007}, if the time intervals overlap, there is a danger of artificially smoothing out variations in \psup~which may intrinsically be discontinuous. Secondly, the choice of the length $\tau$~of the time-intervals is not immediately obvious. At least a few cycles per interval are required to detect cyclic variation, but if $\tau$~is too great then the sensitivity to periodicity variation is lost.

We therefore characterized the DPS of SMC X-1 at two levels. The first investigation broke the total time baseline into 25 {\it non-}overlapping intervals, allowing the uncertainty in period detection to be estimated for a particular strip-length $\tau$~(Section \ref{sssec:rednoise}). For the second investigation, the behaviour of the DPS was charted for a wide range of $\tau$~values (Section \ref{sssec:deltat}).

\subsubsection{Period uncertainty for a given strip-length}\label{sssec:rednoise}

The detection of stable periodicities in RXTE/ASM data is not always straightforward, as the periodogram shows structure due to windowing from the uneven sampling leading to structures in the periodogram which might mimic periodicities or shift the position of detected peaks \citep[e.g.][]{scargle82}. Additionally, the underlying noise structure can often include timescale-dependent ``red noise" that increases the false alarm probability for a periodic signal at a given LS power \citep[e.g.][]{homer2001}. 

We therefore conducted a suite of Monte Carlo simulations to assess the likely period detection uncertainty associated with the DPS approach, tuning the parameters of the underlying noise model and injected periodicities to reproduce the observed peaks, the fine structure, and the underlying slope and plateaux of the LS power spectrum of the true data. The total time baseline was broken into 25 non-overlapping strips of equal duration (at $\tau = 230$~days per strip, this allows 3-4 superorbital cycles per strip; e.g. Figure \ref{fig:lcurve}).

Full details about the determination of superorbital period measurement uncertainty are presented in Appendix \ref{app:sec:uncty}. Briefly, the synthetic lightcurves were generated by adding a  smoothed representation of the detected superorbital modulation to a signal-free time-series with timescale-dependent red-noise following the methodology of \citet{timmerKoenig1995} as implemented in the Python module {\tt DELCgen.py} \citep{connollyDELC}.\footnote{\url{https://github.com/samconnolly/DELightcurveSimulation}} In tests, we found strong variation in execution time between the strips; a procedure to remove this variation, which afforded a factor $\sim 20$~reduction in total simulation time, is communicated in Appendix \ref{ap:timing}.

\begin{figure}
\begin{center}
      \includegraphics[width=3.5in]{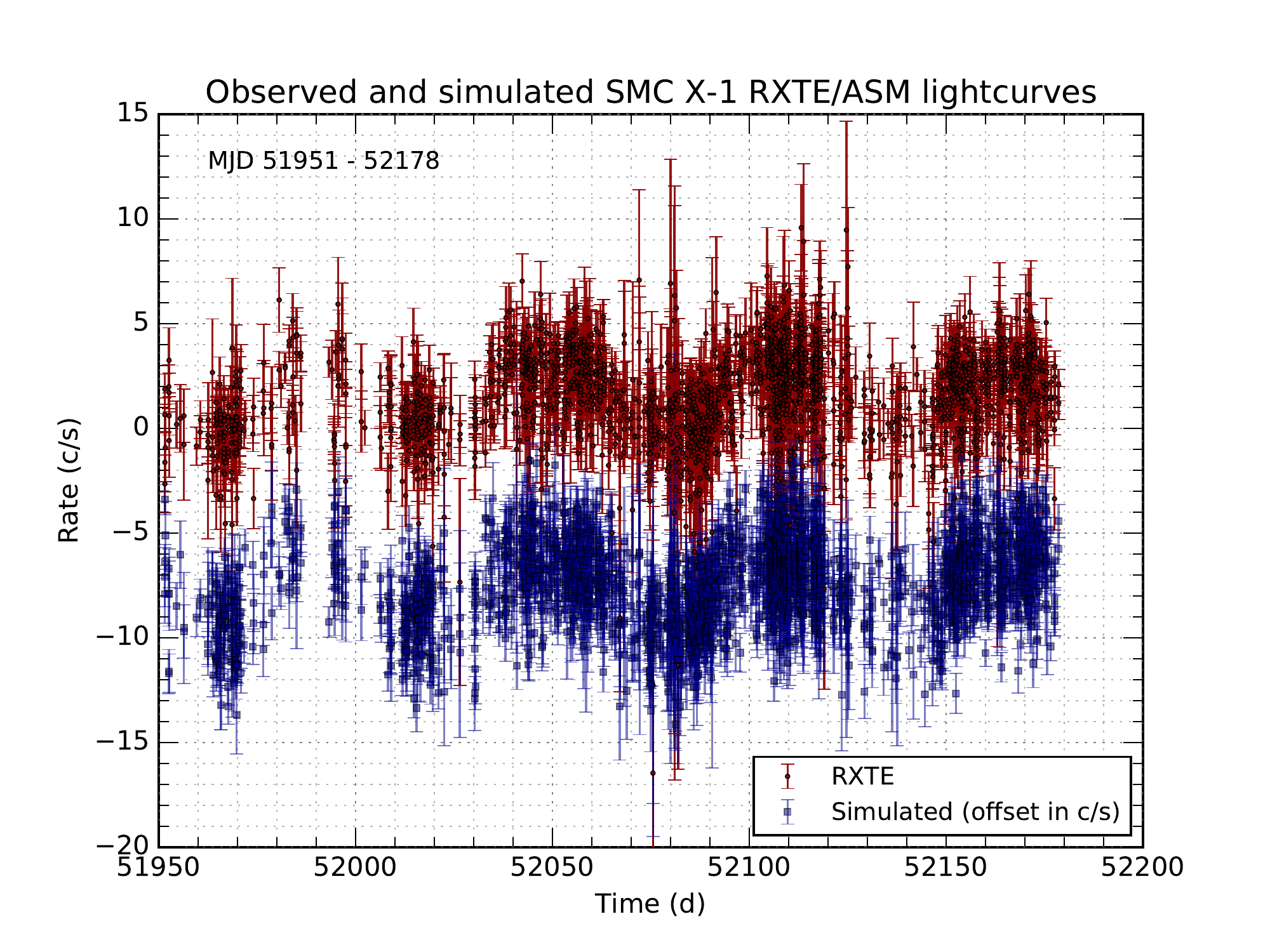}
      \includegraphics[width=3.5in]{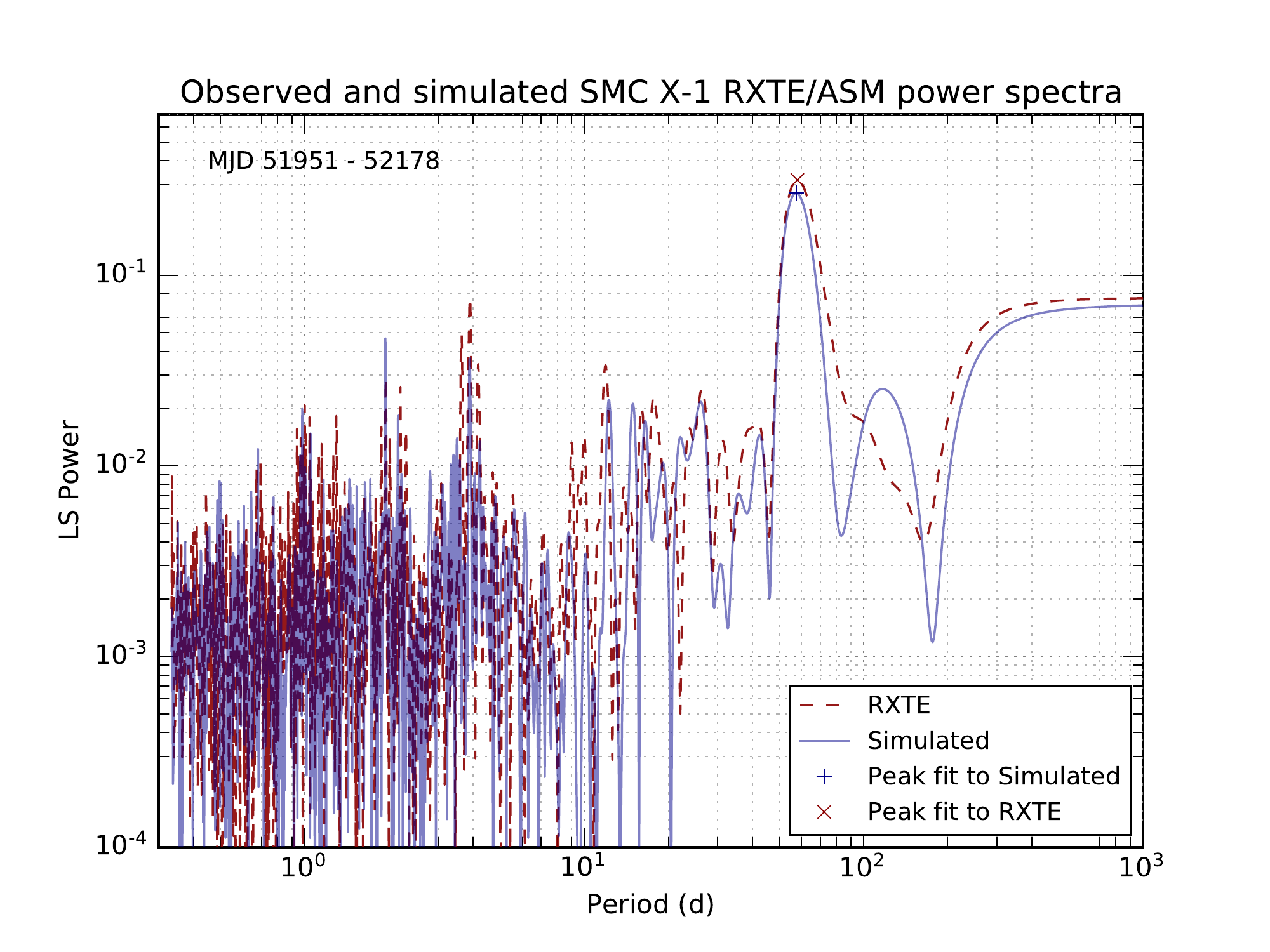}
\caption{{\it Top:} Lightcurve for a data strip of length $\tau \approx 230$~days (MJD $\approx$~51,950-52,180); the true RXTE/ASM is shown in red, the simulated lightcurve in blue and offset for clarity. {\it Bottom:} Log-log power spectrum of both RXTE data and simulated data showing the superborbital variation, as well as the 3.89-day orbit, over a timescale-dependent noise continuum. Both the structure due to red-noise and the uneven sampling are clear. See Section \ref{sssec:rednoise}  and Appendix \ref{app:sec:uncty}.}
\label{fig:cutthru}
\end{center}
\end{figure}

The resulting synthetic lightcurves resemble the observed ASM dataset quite well (Figure \ref{fig:cutthru} shows an example lightcurve and power spectrum for one of the synthetic lightcurves compared to that measured for SMC X-1). We were unable to reproduce the observed periodogram structure using a pure-noise model (with no superorbital periodicity added); a pure-noise model with break frequency at or near 1/\psup~and amplitude tuned to reproduce  observed superorbital modulation, produces too much power (by a factor 10 or more) at timescales $> \psup$.  

\begin{figure*}
\begin{center}
\includegraphics[width=6in]{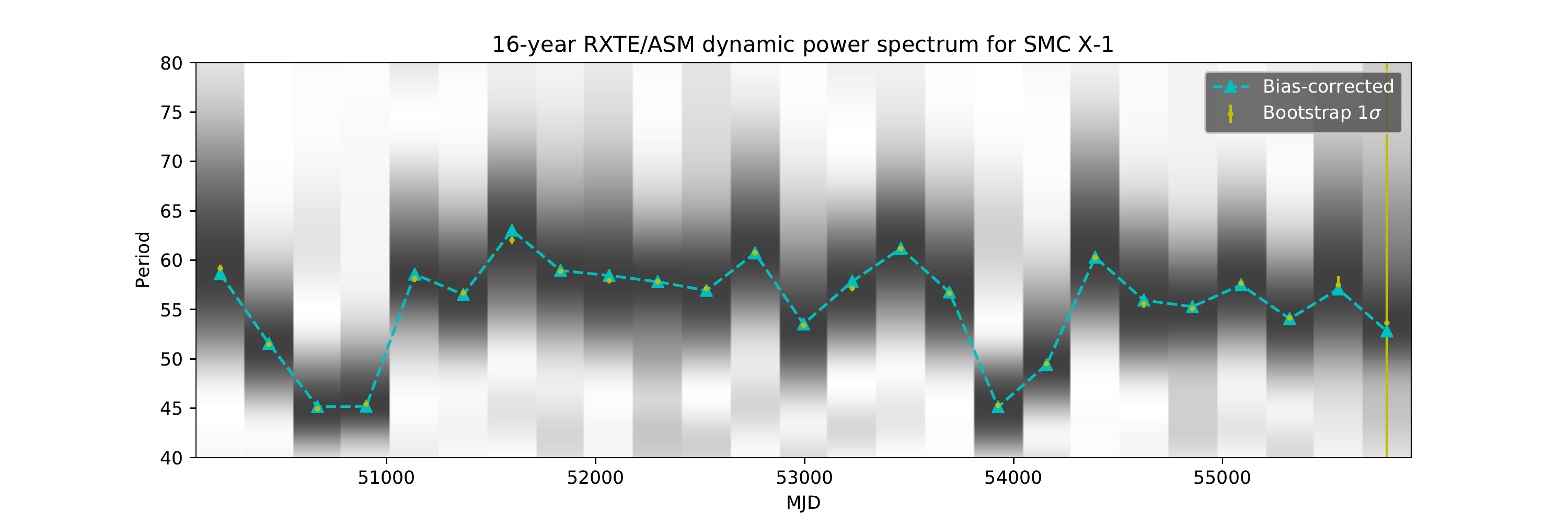}
\includegraphics[width=6in]{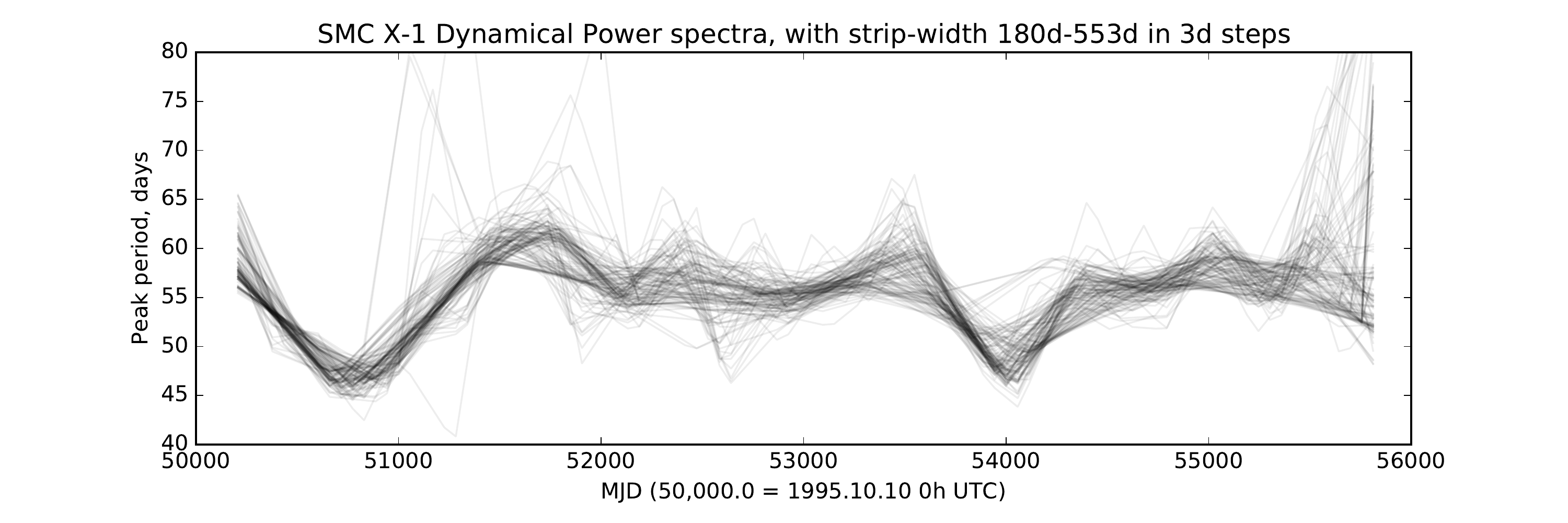}
\caption{Two estimates of the evolving superorbital period \psup~from RXTE/ASM data, using forms of the Dynamical Power Spectrum (DPS). {\it Top:} The time-series is broken into 25 non-overlapping strips of equal length, and the Lomb-Scargle periodogram computed separately for each. The grayscale shows the DPS, with each strip renormalized to a maximum LS power of 1.0. Yellow errorbars are centered on the detected peak from each strip, and represent the $1\sigma$~uncertainties from Monte Carlo trials (see Section \ref{sssec:rednoise} for details). The Cyan triangles show the peak periods, offset by the median bias (in the sense detected minus simulated) suggested by Monte Carlo trials. {\it Bottom:} Maxima in the dynamical power spectrum, where the total $\sim$5600-day time interval is broken into non-overlapping strips of duration $\tau$; the complete run of DPS for $(180$~d $\le \tau \le 553$d$)$~is shown, in steps of $\tau$~of three days. See Section \ref{ssec:psup}.}
\label{fig:superorbs}
\end{center}
\end{figure*}

The top panel of Figure \ref{fig:superorbs} shows the detected superorbital periods and the estimated measurement uncertainties (since these uncertainties are small, see also Figure \ref{fig:app:periodUncty} in Appendix \ref{app:sec:uncty}). All but the final, poorly-sampled strip show random uncertainty in recovered superorbital period in the range $0.22 \le {\rm rms}(\psup) \le 0.44$~d, with a high outlier at $0.90$~days. The detections also show a small bias, in the sense $-1.0 \le$~(measured - simulated) $\le +0.5$~d, although for the majority of strips this bias is in the range $\pm 0.25$~days.

From our Monte Carlo trials we therefore conclude the following: firstly, random uncertainty due to the Lomb-Scargle measurement method is not a significant contributor to the uncertainty in detected \psup; secondly, it is highly unlikely that the superorbital excursions near MJD 50800 and 54000 are spurious.

\subsubsection{Allowing the strip-length to vary}\label{sssec:deltat}

Use of the DPS forces a choice as to the strip length $\tau$. A particular choice of $\tau$~might cause intrinsic variation to be missed or artificially enhanced, depending on where the strip boundary lies with respect to intrinsic variation in either the profile or the period of superorbital modulation (see also \citealt{trowbridge2007}).

As a second method to estimate uncertainty in \psup, we therefore computed the DPS for a wide range of strip-lengths $\tau$~and examined the ensemble of DPS estimates thus produced for commonalities in structure.  The uncertainty estimate thus produced contains contributions from intrinsic variation in lightcurve structure, from statistical uncertainty due to both measurement uncertainty and temporal sampling variations, and from the range of choices of strip-length.

Where the detected period near a given date shows a small range of values over this ensemble, we interpret this as suggesting variation that is intrinsically reasonably stable. Conversely, where the detected period shows a large range for a particular date, we interpret this as indicating intrinsic changes on a short enough timescale that the choice of $\tau$~does impact the recovered value of \psup.

The bottom panel of Figure \ref{fig:superorbs} shows the evolution of the superorbital peak in the DPS as computed separately for 3-day steps in $180 \le \tau \le 553$~days. The first and second superorbital excursions near MJD 50,800 and MJD 54,000 are visible with all strip-lengths, although the dates of the period minima can change by a few months depending on the value of $\tau$. (Smoothness in the \psup~curves may be illusory, since each value of $\tau$~represents averaging over a different number of superorbital cycles.)

The scatter in recovered \psup~for all dates is larger (by about a factor 5-10) than that suggested by the Monte Carlo trials with constant \psup~(Section \ref{sssec:rednoise}). We therefore adopt the rms scatter over the ensemble of strip-lengths $\tau$~as our running estimate of superorbital period uncertainty for the rest of this work.

\subsection{Changes in spin period}\label{ssec:pspin}

\begin{figure*}
\begin{center}
\includegraphics[width=18cm]{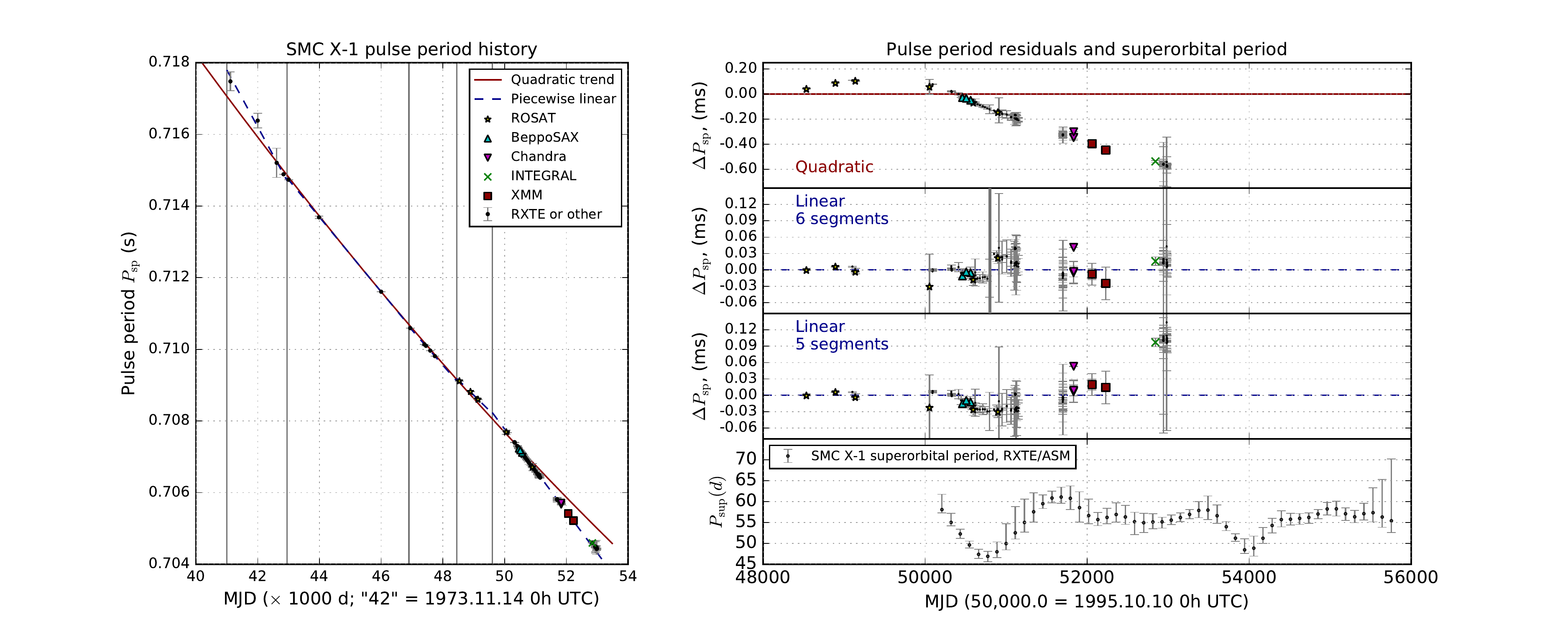}
\caption{ Pulse period history of SMC X-1, from the compilation of \citetalias{inam2010}, supplemented by pulse periods reported in \citetalias{neilsen04}. {\it Left panel:} the pulse period history. {\it Right panels:} residuals after a quadratic (right-top) and piecewise linear trends (right upper- and lower-middle panels.) The piecewise linear fit with five segments uses the same time partitioning as \citetalias{inam2010}, which assigns the entire \rxte~mission to a single partition. For the six-segment scheme, the \rxte~era is partitioned at MJD 50,800 days (the vertical gray line in the the right upper-middle panel). Plot symbols and colors indicate different missions: after MJD 48,000, black points with errorbars indicate \rxte~observations, while before this date the same symbols indicate ten missions ({\it Uhuru, Aerobee, Apollo-Soyuz, SAS 3, Ariel V, Einstein, EXOSAT, Ginga, HEXE} or {\it ASCA}). In all panels, the dark red solid line indicates the quadratic trend fit to the entire pulse period history, while the dark blue dashed line shows the piecewise linear fits. Finally, the {\it right-bottom} panel shows the \rxteasm~superorbital period determination; medians are interpolated from the run over lightcurve strip-lengths $\tau$~presented in the bottom panel of Figure \ref{fig:superorbs}, with  asymmetric $1\sigma$~errorbars indicating the root mean square of the variation of superorbital period above and below the median detected values. See Section \ref{ssec:pspin}.}
\label{fig:pPulse}
\label{fig:pPulseSix}
\end{center}
\end{figure*}

Figure \ref{fig:pPulse} shows the pulse period history of SMC X-1. As shown by \citetalias{inam2010}, SMC X-1 exhibits a long-term spin-up trend, consistent in sense (if not magnitude) over the almost five decades for which this source was observed.\footnote{SMC X-1's pulse spin-up is not entirely monotonic; for example, a short spin-down event is visible in the pulse periods in the interval MJD 51,100-51,125 - itself during an interval with unusually intensive time-sampling.} As pointed out by \citetalias{inam2010}, the magnitude of this spin-up changes on a timescale of years. \citetalias{inam2010} divide their four-decade dataset into five segments, each separately fit with a linear trend (some of the segments have fewer than four samples in each). For the \rxte~era the pulse history becomes substantially better-sampled, with $>$100 pulse-period measurements in the MJD 50,000-53,000 interval. We followed \citetalias{inam2010} and fit a piecewise linear trend to the pulse period history, adopting the same partitioning by time interval.  When fitting the spin-period trends to estimate \pspin, the interval MJD $>$52,500 was explicitly avoided in the fitting to avoid artificially pulling this time interval back to the baseline behaviour. (For completeness, we did attempt to fit a single low-order polynomial to the entire pulse-period history of SMC X-1 in the manner of \citet{woj98}, but found that this process invariably left discontinuities in the pulse period history when the complete time range of \citetalias{inam2010} was considered.)

\subsubsection{Estimating the smoothed pulse period derivative}\label{ss:pderiv}

Under our working hypothesis, if the spin- and superorbital variations really are related to each other, then we should see changes in \pspin~ on the same range of timescales as those in \psup. We are therefore interested in variations in \pspin~on the order of weeks-years. That \citetalias{inam2010} \& \citetalias{neilsen04} do not report the {\it instantaneous} \pspin~derivatives is therefore of no real consequence to the present investigation.

To isolate changes in  spin-period derivative on timescales of interest, the detrended values of \pspin~were binned to 50 days (approximately one superorbital cycle) and the period derivative estimated as the derivative of a spline fit to the resulting smoothed periods. Use of this quantity, \dpspin, is analogous to the application of a low-pass filter tuned to the average superorbital cycle length. 

\section{Results}\label{sec:results}

The comparison of superorbital period \psup~with the pulse period \pspin~and the smoothed pulse period gradient \dpspin~is shown in Figures \ref{fig:pPulse} and \ref{fig:overplot}. If there is a strict correlation between \psup~and either \psup~or \dpspin, it is not simple or stable. 

\subsection{Variation of \texorpdfstring{\pspin}{spin period}~with \texorpdfstring{\psup}{superorbital period} }\label{ss:resPspin}

The first superorbital excursion (at MJD $\approx 50500-51500$) coincides with an apparent shortening in \pspin~over about the same timescale. However the \pspin~behaviour after this interval does not closely resemble the changes in \psup. Although coverage is sparse near MJD 53,000, the \pspin~measurements in that region show no sign of returning to a baseline level, as \psup~appears to do (Figure \ref{fig:pPulse}). 

\subsection{Variation of \texorpdfstring{\dpspin}{smoothed spin period gradient}~with \texorpdfstring{\psup}{superorbital period}}\label{ss:resPdot}

\begin{figure}
\begin{center}
\includegraphics[width=8cm]{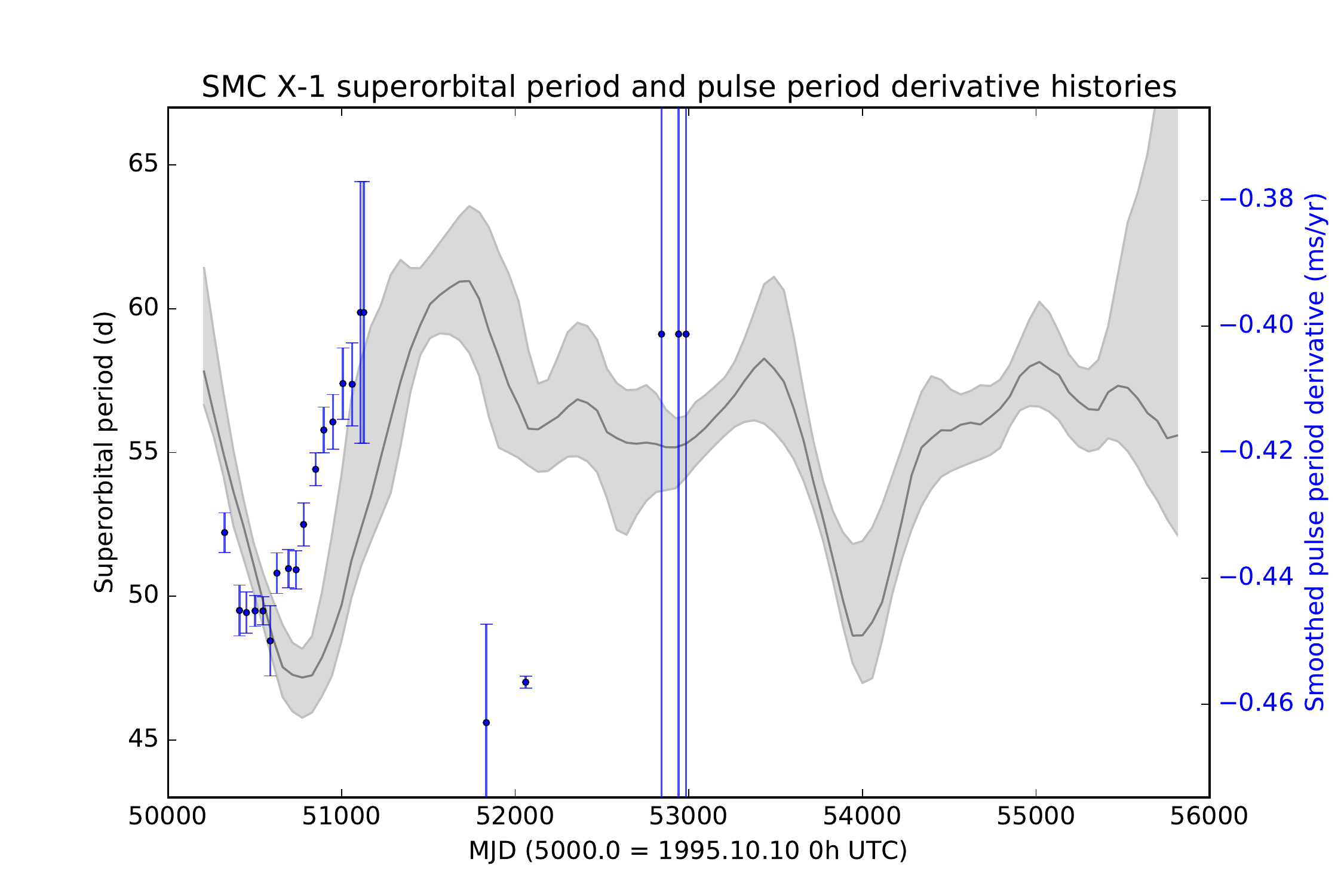}
\caption{Smoothed pulse period derivative history overlaid on the superorbital variations.The superorbital period variation is represented by the median values of \psup~over the range of $\tau$~values, with 1$\sigma$~upper and lower bounds indicated as grayscale. See Section \ref{sec:results}.}
\label{fig:overplot}
\end{center}
\end{figure}

Near the first superorbital excursion (MJD $\approx 50,800$), the smoothed \dpspin~history shows a striking similarity to the behaviour of \psup, with  changes in \dpspin~taking place about 200 days before the changes in \psup~(Figure \ref{fig:overplot}). As with the detrended spin period \pspin, however, the intervals at MJD $>$51,500 do not obviously confirm the trend - although we caution again that the coverage in pulse period measurements is relatively sparse in this interval.

\section{Discussion}\label{sec:discussion}

Even with the huge \pspin~dataset amassed by \citetalias{inam2010}, complemented by the measurements of \citetalias{neilsen04}, we cannot claim a detection of a strict correlation between \psup~and either \pspin~or \dpspin. Both these quantities do show variations that appear to {\it coincide} with variations in \psup, but only for the interval near the first superorbital excursion at MJD $\approx$~50,800. 

\subsection{X-ray modulation scenarios}

Without coverage of the {\it second} superorbital period excursion near MJD 54,000, a number of interpretations are still consistent with the publicly-available RXTE measurements. We consider the alternatives in turn.

\subsubsection{Correlated excursions in both \texorpdfstring{\pspin}{spin}~and \texorpdfstring{\psup}{superorbital period}}\label{ss:dis:pspin}

One possibility is that \psup~and \pspin~both experience an excursion to shorter periods nearly simultaneously. This might indicate a scenario in which the accretion disc and inner accretion flow both change simultaneously. However, behaviour after the first \psup~excursion is markedly different, and, under this scenario, it is not clear why the cluster of \pspin~measurements near MJD 53,000 does not follow the same behaviour.

\subsubsection{A change in slope of \texorpdfstring{\pspin}{spin period}, coincident with an excursion in \texorpdfstring{\psup}{superorbital period}}\label{ss:dis:torque}

As shown by \citetalias[][and refs therein]{inam2010}, and as we confirm for this study, SMC X-1's pulse period history is best characterised as a series of linear segments of differing gradient but all decreasing in period, rather than a single low-order polynomial over its entire spin period history (Figure \ref{fig:pPulse}). Discontinuous changes in spin period derivative are not unknown in X-ray pulsars; examples include GX 1+4 \citep{chakrabarty97}, 4U 1626-67 \citep[which, like SMC X-1, is Roche-lobe fed]{camero-arranz10} and GX 301-2 \citep[e.g.][]{kaper06}, although in those cases the changes in spin period are much more dramatic than in SMC X-1, which never undergoes an extended interval of torque reversal. Models to explain these changes in spin period gradient generally suggest a change in accretion rate onto the pulsar \citep[e.g.][]{yiVishniac99}, possibly caused by changes in the inner accretion disc \citep[e.g.][]{wijers1999}. For the case of SMC X-1, the discontinuous changes in \pdot~have generally taken place between intervals of
\pspin~coverage  \citepalias[][ also the left panel of Figure \ref{fig:pPulse}]{inam2010}. 

However, it may be that what we interpreted in Section \ref{ss:dis:pspin} as an excursion in \pspin~near MJD 50,800, may more accurately be described as another of the discontinuous changes in  the spin period derivative \pdot. Under this interpretation, the spin period behaviour near MJD 50,800 may represent a transition to a slightly slower spin-up rate from the SMC X-1 pulsar. The right-hand panel of Figure \ref{fig:pPulseSix} shows pulse period residuals over the RXTE era, divided at MJD 50,800, approximately the epoch of the first superorbital excursion. Under this model, the spin period gradient changes from $d\pspin / dt$ = -0.432(2)~ms yr$^{-1}$~before MJD 50,800 to -0.396(6)~ms yr$^{-1}$~after it (with the figures in parentheses giving the formal uncertainty in the last decimal place; these gradients are consistent with the smoothed period gradients shown in Figure \ref{fig:overplot}). The reduced chi-square statistic $\chi^2_\nu$~is unlikely to show values close to unity when linear trends are used to fit structure as complicated as that in SMC X-1's pulse period history (particularly in the presence of a small number of measurements with extremely small uncertainty estimate; \citetalias{inam2010}), but we do see an improvement in the fit when the RXTE era is partitioned in this way: representing the RXTE era with a single linear trend produces $\chi^2_\nu = 122,738/135 = 909.2$~while a two-segment fit produces $\chi^2_{\nu} = 45,244/133 = 340.2$. Thus, while these $\chi^2_\nu$~are far from formally acceptable,\footnote{Under the hypothesis that residual variation after subtraction of the best-fit model is due to measurement uncertainty, producing $\chi^2_\nu \approx 1$~would require multiplying the pulse period uncertainties by a factor $\sim 18$, which seems unreasonably large.} they do suggest improvement in the fit to the pulse period history when a two-segment fit is used for the RXTE era. 

We thus suspect that the compilation of \citetalias{inam2010} may have caught a torque-change event from SMC X-1 near MJD$\approx 50,800$. If so, then the \psup~excursion near MJD 50,800 may relate to the putative torque change near the same date. With these data we are unable to match the events more accurately than a few viscous timescales. 

\subsubsection{A link between \texorpdfstring{\dpspin}{smoothed spin period derivative}-and \texorpdfstring{\psup}{superorbital period}?}\label{ss:dis:dpspin}

If indeed \dpspin~and \psup~are varying together, a time-lag of $\Delta t \sim 200$~days between variations in \psup~and \dpspin~might be expected. If the smoothed \dpspin~traces changes in the accretion rate onto the neutron star, then the interception of accretion luminosity by the warped disc might lead to adjustment in the disc configuration. With disc adjustments taking place on one to a few viscous timescales \citepalias{od01}, this might then lead to changes in measured \psup~that lag those in \dpspin~by $\sim 200$~days.

\subsubsection{No relation between the spin- and superorbital behaviours}\label{ss:dis:nocorr}

A final possibility is that there is no intrinsic link between the variation in superorbital and pulse period. We note that the data considered in this paper together do {\it not} rule out this possibility. In this scenario, the accretion disc and the pulsar both do respond to local changes in accretion rate, but changes in the accretion flow are so complex that variations in the accretion disc configuration are observationally decoupled from variations in the pulse period. We point out that this scenario would not invalidate the radiation-driven warping scenario (in which the accretion luminosity from the accretor drives the disc into its warped configuration; \citealt{wijers1999}, \citetalias{od01}), but instead the changes in the pulsar and the disc are too intrinsically noisy to correlate.

\subsection{Pointed X-ray coverage of the second \texorpdfstring{\psup}{superorbital period}~excursion}\label{ss:dis:moreObs}

The second superorbital excursion, at approximately MJD 53,500-54,500 (\citealt{trowbridge2007, kotzeCharles2012}, roughly the 2005 May - 2008 Feb interval) is not covered in the \citetalias{inam2010} compendium of pulse periods. We examine here the archival coverage of pointed observations which may allow \pspin~measurement during and after the second superorbital excursion.

To our knowledge, the most recent pointed RXTE/PCA measurements sensitive to the pulse period, are those obtained on MJD 52,981-52,987 (2003 Dec 08-14) and published in \citet{raichurPaul10} and \citet{falanga15}. A recent examination of type-II bursts from SMC X-1 in its high state also uses RXTE/PCA data no more recent than 2003 December; \citet{rai18}. 

The HEASARC archive shows only a single set of RXTE/PCA pointed observations with coverage beyond 2003 December. This is RXTE program 80078 (PI Eikenberry), which observed SMC X-1 several times in the 2003 October - 2004 January interval. A subset of these observations were used by \citetalias{inam2010} and are tabulated in their Table 3. Curiously, the 2004 January observations within program 80078 do not appear in \citetalias{inam2010}, and a subsequent processing using the Southampton RXTE/PCA pipeline (see \citealt{Yang2017})\footnote{\url{http://www.xraypulsars.space/projects}} confirms that the pulsar is not detected in those observations. These observations did not take place during mid-eclipse (using the ephemeris of \citealt{falanga15}),  nor precisely at superorbital minimum, and the RXTE/ASM did detect SMC X-1 over the same interval, so it remains unclear why the SMC X-1 pulsar has not been detected in the 2004 January observations. Regardless, for dates more recent than 2004 January, the HEASARC archive does not contain any RXTE/PCA pointings within one degree of SMC X-1 (roughly the distance off-axis within which RXTE/PCA has sensitivity). 

While SMC X-1 has been observed several times by other missions during and after the second superorbital excursion (roughly 2005 May - 2008 Feb), it is not immediately obvious that any of the measurements will allow useful measurements of SMC X-1's pulse period. For example, SMC X-1 was within the {\it ACIS-I} field of view for \chandra~observations 5493 \& 5494 (both P.I. Coe, taken during 2008 Feb and March), but the 3.2s time resolution of {\it ACIS-I} is too poor to permit pulse period measurement of this source. Of the five {\it Swift} observations made of SMC X-1 during the 2006-2009 interval (obsIDs 00035216001, 00035216002, 00035216003, 00037160002 \& 00037160003), only the first was taken in a mode with sufficient time resolution to measure SMC X-1's spin period. However, as these data were taken in Windowed-Timing mode (by which the image is collapsed along one spatial axis) and SMC X-1 itself was not at the aim-point of the observation, it is not clear that SMC X-1's spin period can be measured from these data. Finally, {\it XMM} observed the Nova SMC 2005 field, including SMC X-1, during 2006 March 13th (MJD $53,807.64 - 53,807.77$, ObsID 0311590601, PI Schartel). Using the ephemeris presented in Table 3 of \citet{falanga15}, this observation took place at orbital phase $0.90-0.94$. With eclipse commencing at phase $\approx 0.90$~\citep{falanga15}, it is likely that extracting a pulse period measurement from this latter observation may also be problematic.

 Extraction of pulse periods from this heterogeneous and patchy archival dataset near the second superorbital excursion is beyond the scope of the present investigation; at this juncture we expect perhaps 0-2 pulse period measurements will be extractable from this interval. However, the {\it MAXI}~satellite continues to furnish long-term X-ray monitoring coverage of bright X-ray binaries in a similar manner to the \rxteasm~\citep[e.g.][]{sugimoto16}, including SMC X-1 itself \citep{sugimoto14}. Therefore, we anticipate X-ray monitoring of further superorbital period excursions of SMC X-1 will allow triggering of observational campaigns to measure \pspin~evolution coincident with further large changes in \psup.

\subsection{Predictions for future \texorpdfstring{\psup}{superorbital period}~excursions}\label{ss:dis:pred}

While comprehensive mining of archival data for MJD $\ge$~53,000 is beyond the scope of this paper, we can make the following predictions for what analysis of the full archive will reveal, depending on the scenario:
\begin{itemize}
\item{If the \pspin~and \psup~excursions are strictly correlated, then \pspin~will decrease during  the second and future superorbital period excursions, returning to a baseline level  after the end of the excursion (Section \ref{ss:dis:pspin});}
\item{If instead the \psup~excursion heralds a sudden change in torque, then the spin period gradient \pdot~will change near  the middle of the excursion, and remain roughly constant until any future superorbital excursions take place (Section \ref{ss:dis:torque});} 
\item{If the smoothed spin period gradient \dpspin~varies with \psup, then the variation of this quantity will resemble  that of the \psup~excursion itself, just shifted earlier by $\sim 200$~days (Section \ref{ss:dis:dpspin});}
\item{If instead there is no link between \pspin~and \psup, then the \pspin~behaviour during superorbital excursion will show no relation to that of \psup~(Section \ref{ss:dis:nocorr}).}
\end{itemize}

\subsection{Comparison to Her X-1}\label{ss:herx1}

 The eclipsing X-ray binary Her X-1 (\pspin~1.24s, orbital period 1.70d; donor mass 2.2 $M_{\odot}$, e.g. \citealt{jurua2011}) provides a second example of an X-ray binary system in which both the X-ray pulse behaviour and superorbital modulation may be tracked together, providing important context for the present study. This source shows a 35-day superorbital modulation, with intrinsic scatter of about $\pm 2$~days (e.g. Figure 5 of \citealt{leahy2010a}), or about $\pm 7\%$~variation in \psup~compared to $\pm 20\%$~for SMC X-1. The relative stability of the superorbital period exhibited by Her X-1 is broadly consistent with its system properties under the radiation-driven warping framework (e.g. \citetalias{od01}; \citetalias{clarkson2003b}). Under this framework, the figure of the accretion disc warp in Her X-1 thus likely changes less severely than is the case for SMC X-1, but is still characterized by a range of superorbital periods and associated disc configurations against which other behaviours can be compared. A correlation has been measured between \psup~and the average X-ray flux during a superorbital cycle \citep{leahy2010a}, which might suggest a link between \psup~and $\dot{M}$.

 The relatively stable superorbital modulation of Her X-1 is complicated by occasional ``Anomalous Low States" (ALS), in which the observed X-ray flux drops to very faint levels for intervals of a few months, coincident with and possibly also preceded by intervals of rapid spin-down (e.g. \citealt{parmar1999}, \citealt{staubert2006}) and most likely corresponding to an increase in X-ray obscuration rather than a cessation of $\dot{M}$~(see \citealt{still2001} and references therein). The currently favored model is that the ALS state indicates a small change in the form of the accretion disc warp, leading to both an increase in X-ray obscuration and altering the instantaneous $\dot{M}$~through the inner disc \citep[e.g.][]{staubert2006}. That the superorbital modulation at {\it optical} wavelengths shows very little change between the ALS and normal high-state, suggests that the disc warp need not change strongly in its form to produce the large increase in X-ray obscuration characterizing the ALS \citep[][and references therein]{jurua2011}.

 Outside ALS events, Her X-1 shows strong correlation between the pulse period and the phasing of the main-on in the superorbital cycle, with the superorbital main-on variations lagging behind the \pspin~variations by a few tens of days \citep{staubert2006}. Additionally, variations in the pulse profile from Her X-1 are precisely synchronised with superorbital main-on phasing \citep{staubert2013}. The two leading candidates to account for the pulse profile variations from Her X-1 are free precession of the neutron star (e.g. \citealt{postnov2013} and references therein), and varying occultation of the neutron star by the inner accretion disk \citep{scott2000}, although both models have difficulties (see \citealt{staubert2013} for discussion).

 The two sources differ in the energy dependence of pulse profile variations. From Her X-1, the pulse profile at $9-13$~keV is observed to vary \citep[e.g.][]{staubert2013}, while from SMC X-1 it is only the soft (blackbody, at $\lesssim 1.0$~keV) component of the pulse profile that varies with superorbital phase \citep{hickox2005}. While the \citet{scott2000} mechanism invokes progressive occultation by the inner accretion disk to explain pulse profile variations seen from Her X-1, the \citet{hickox2005} mechanism invokes varying reprocessing at the inner disk to explain pulse profile variations seen from SMC X-1.

 While we have not performed precisely the same tests as \citet{staubert2006} - we characterize the period of the superorbital modulation from SMC X-1 over several cycles rather than its phasing from individual cycles - our results from SMC X-1 are probably intermediate between the lockstep correlation observed from Her X-1 and a case with no correlation at all. We see an excursion in \pspin~that appears to coincide with an excursion in \psup~(and, in the case of the smoothed \pspin~derivative, also possibly leading the superorbital modulation by tens of days) but spin-superorbital correlation does not appear to recur later in the \rxte~interval.

 The observed behaviour from SMC X-1 is probably more consistent with ALS-type behaviour (a small change in accretion disc-warp figure accompanied by an episode of pulsar spin-down) than with the correlated changes in (presumed) accretion warp figure and pulse profile seen from Her X-1. While the physics underlying the (presumed) response of the accretion disc to changes in the pulsar behaviour may be expected to be similar for SMC X-1 as for Her X-1 (with the main difference being the stability of the accretion disc warp configuration and the complexity of the form of the warp), more detailed cross-comparison of the two systems probably awaits X-ray pulse period measurements during another superorbital excursion from SMC X-1.

\section{Conclusions}\label{sec:conclusions}

With X-ray variability measurements that trace both superorbital and pulsar spin-period timescales,  and with large-amplitude ($\sim 20\%$) variation in superorbital period, SMC X-1 is perhaps the best test-case for probing the interplay between the behaviour of a large, persistent accretion disc, and variations in the accretion flow onto the compact object. We have therefore performed an investigation to determine whether variations in the pulsar spin period derivative are correlated with long-term variations in the superorbital X-ray variability of this source. Our conclusions are:

\begin{itemize}
  \item{We find an intriguing possible correlation between the behaviour of \psup~and the pulsar spin period, particularly in the interval around the first \psup~excursion at MJD $\approx$50800;}
    \item{The measurements do not yet allow us to discriminate between several possible scenarios. The measurements are most consistent with either a model in which the accretion disc and pulsar are both responding to a discontinuous change in accretion flow, possibly regulated by the disc (Section \ref{ss:dis:torque}), or one in which the variations in pulsar torque and accretion disc configuration are too complicated to be observationally coupled (Section \ref{ss:dis:nocorr});}
\item{The {\it second} \psup~excursion near MJD 54,000 offers an opportunity to distinguish between these scenarios,  at least in principle, but the coverage of observations {\it sufficient to measure the pulse period} is likely much more sparse, and we defer detailed analysis of these observations to further work;}
\item{ With ongoing long-term X-ray monitoring from {\it MAXI}, future superorbital period excursions may allow renewed observational probing of the variation of SMC X-1's spin period evolution as the superorbital period changes.}
\end{itemize}

We therefore believe that a follow-up to \citetalias{inam2010} is now required, to assess the full pulse-period history of SMC X-1. By comparing the pulse period history during the complete RXTE interval to the superorbital modulation, including the second superorbital excursion near MJD 54,000 and subsequent recovery, it may be possible to reveal if and how the accretion disc and pulsar really are responding to the same underlying processes.  If, however, archival observations are indeed not sufficient to probe the spin period evolution during this second superorbital excursion, then further probing of the spin-superorbital link in SMC X-1 may have to await future observations, triggered by long-term monitoring by {\it MAXI} or similar future facilities. This would allow new tests of the radiation-driven warping framework \citepalias{od01}, and finally allow SMC X-1 to yield its secrets into the accretion/outflow process that is so central to astrophysics at all scales.

\section{Acknowledgments} This work was only possible because of the efforts of the ASM/RXTE team at MIT and NASA/GSFC. Data was obtained through the High Energy Astrophysics Science Archive Research Center Online Service, provided by NASA/GSFC. This research made use of {\tt Astropy}, a community-developed core Python
package for Astronomy \citep{astropy2013}. This project made use of the {\tt sympy} symbolic algebra Python library. PAC gratefully acknowledges receipt of a grant from the Leverhulme Trust. SL acknowledges support of NASA Astrophysics Data Analysis program grant 80NSSC18K0430. We thank Malcolm Coe, of the University of Southampton, for insightful discussion. We thank the anonymous referee for a thorough and careful reading of the manuscript, and for thoughtful comments and suggestions which greatly improved the paper. 

\bibliographystyle{mnras}
\bibliography{smcx1_letter_2017W}








\appendix

\section{Lomb-Scargle period detection uncertainty for long-period variability}\label{app:sec:uncty}

Several effects complicate the effort to estimate the measurement uncertainty on the superorbital periods  shown by SMC X-1, as we demonstrate below. With recurrence interval in the range 45-60 days, the superorbital variation from SMC X-1 takes place in a period regime at which timescale-dependent red-noise can be important (e.g. \citealt{homer2001}). Intrinsic variability at these timescales can cause substantial power at other periods, and a period-finding by Gaussian fitting to power spectrum peaks at long timescales can be biased (usually to higher periods) since the peaks themselves are not Gaussian. Pre-treatment of the power spectrum to remove underlying red-noise can circumvent these difficulties to some degree (e.g. \citealt{levine2011}), but to produce superorbital periods and their uncertainties with as few assumptions as possible, we have opted to work with the power spectrum without pre-treatment. This Appendix communicates the techniques used to estimate the superorbital period, the underlying noise model, and thus the measurement uncertainty of the superorbital period for each of the time strips in the \rxteasm~lightcurve.

The superorbital period measurement uncertainty was estimated by generating a large number of synthetic \rxteasm~lightcurves tuned to match the observed lightcurve and power spectrum of each time strip, and the random and systematic components of the period uncertainty estimated by characterizing the distribution of the difference $\Delta P$~between the recovered and simulated superorbital periods. In the process, we accounted for bias due to asymmetry of the power spectrum peak (Appendix \ref{app:ss:peakBias}) and paid particular attention to the shape of the underlying noise continuum in the periodogram (Appendix \ref{app:ss:continuum}).

 For each simulation, the synthetic lightcurve was generated using the \citet{timmerKoenig1995}~method (for a timescale-dependent red-noise continuum) and from unit Gaussian random variates scaled by \rxteasm~measurement uncertainties (testing the white-noise case). Sinusoids at the orbital period and its first harmonic were added to simulate orbital modulation, and the superorbital modulation injected by adding the fold to the observed dataset on the detected superorbital period. In this way, the average superorbital modulation profile for a given strip is preserved, allowing spectral leakage from the superorbital modulation to be accurately included in the simulations. Figure \ref{fig:cutthru} in the main paper shows an example of the quality of the match obtained to the measured data through these methods.

\subsection{Power spectrum peak asymmetry bias}\label{app:ss:peakBias}

\begin{figure}
\includegraphics[width=8cm]{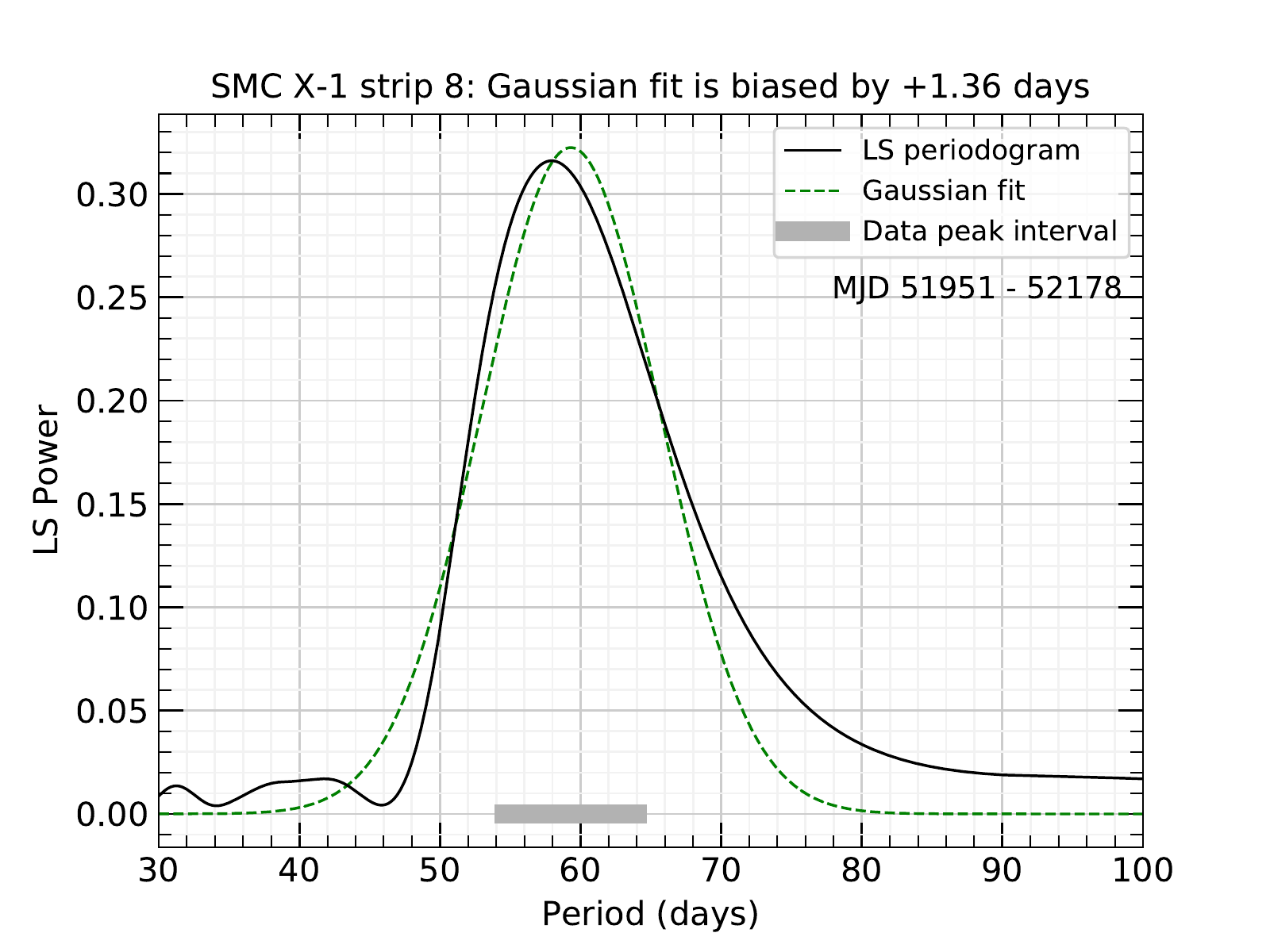}
\caption{ Illustration of bias in a Gaussian fit (dashed line) to the strongest Lomb-Scargle power spectrum peak from SMC X-1 (solid line). Due to asymmetry in the true profile, the Gaussian peak is artificially shifted to higher values. The thick gray bar shows the selection region bracketing the true peak; instead of the peak of the Gaussian approximation, we use the maximum profile value within this selection region to determine the power spectrum peak period. See Section \ref{app:ss:peakBias}.}
\label{fig:app:peakBias}
\end{figure}

At long periods, peaks in the Lomb-Scargle power spectrum from SMC X-1 become sufficiently asymmetric that a Gaussian fit to the peak is biased high by up to about 2.5 days - much larger than the typical random uncertainty we estimate in the period detection. Figure \ref{fig:app:peakBias} shows an example. Instead of using a model fit to the power spectrum peak, we use the Gaussian approximation to determine a region of evaluation that brackets the true peak, then determine the maximum value of the true peak using a highly oversampled evaluation of the power spectrum within this region.

\subsection{Continuum model}\label{app:ss:continuum}

\begin{figure}
\includegraphics[width=8cm]{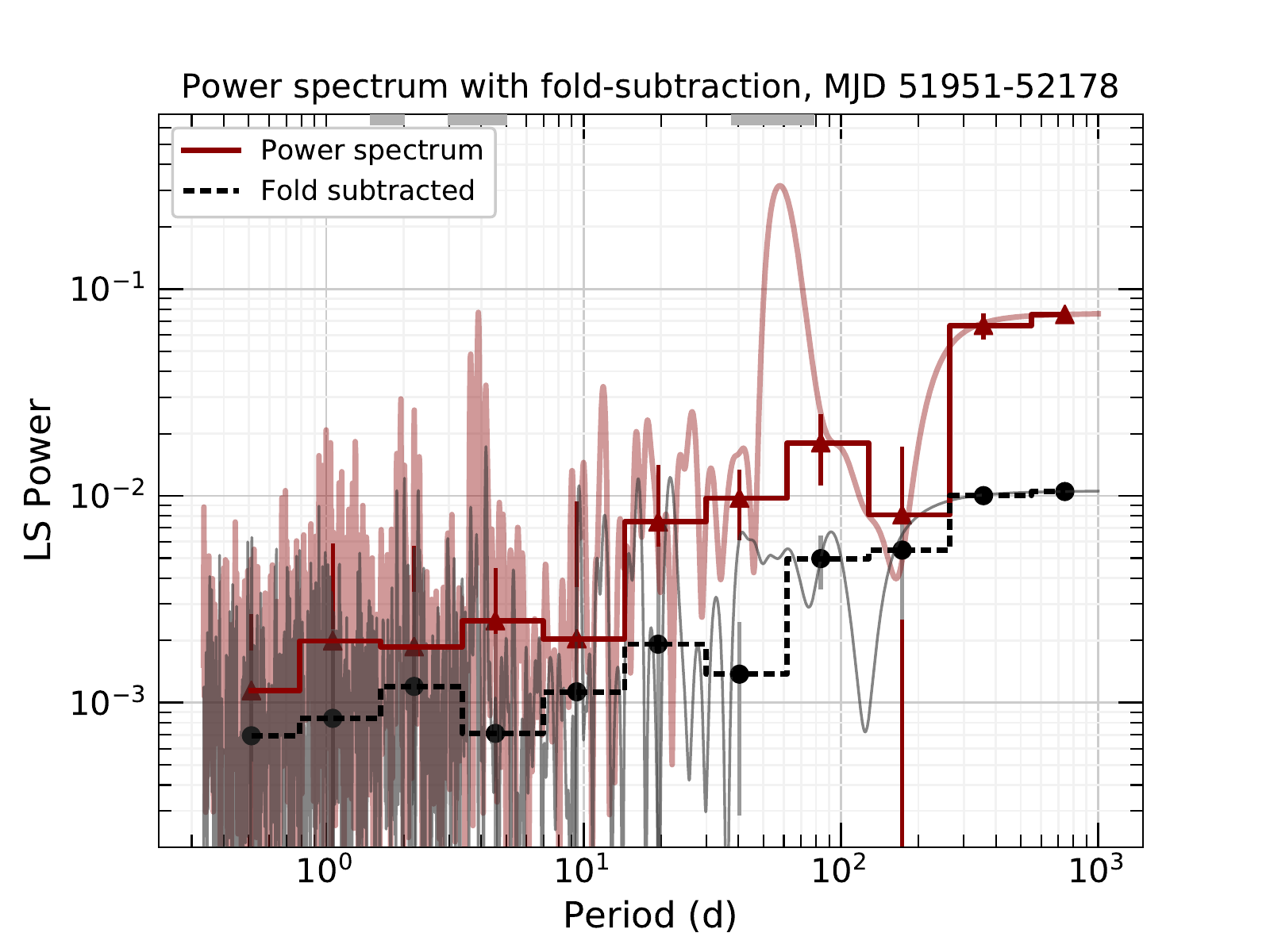}
\caption{ Estimating the power spectrum continuum for a single lightcurve strip. The red curve and step-plot show the observed and binned Lomb-Scargle power spectrum, respectively. The gray curve and dashed black step-plot show the observed and binned power spectrum, after folds on the superorbital and orbital periods are subtracted from the observed lightcurves. The gray bars along the top horizontal axis indicate regions ignored from the continuum fitting due to proximity to the superorbital and orbital power spectrum peaks. See Section \ref{app:ss:continuum}.}
\label{fig:app:continuumObs}
\end{figure}

To characterise the power spectrum continuum, the observed power spectrum was coarsely binned in log-log space (ignoring the period interval surrounding the detected superorbital peak, as well as intervals bracketing the orbital period and its first harmonic). To remove spectral leakage from the superorbital modulation and thus isolate the underlying continuum, a fold on the superorbital modulation was subtracted from the lightcurve, followed by subtraction of a fold on the orbital period. Figure \ref{fig:app:continuumObs} shows an example; the subtraction of measured orbital and superorbital variability reveals an underlying continuum that depends on modulation timescale.

\begin{figure}
\includegraphics[width=8cm]{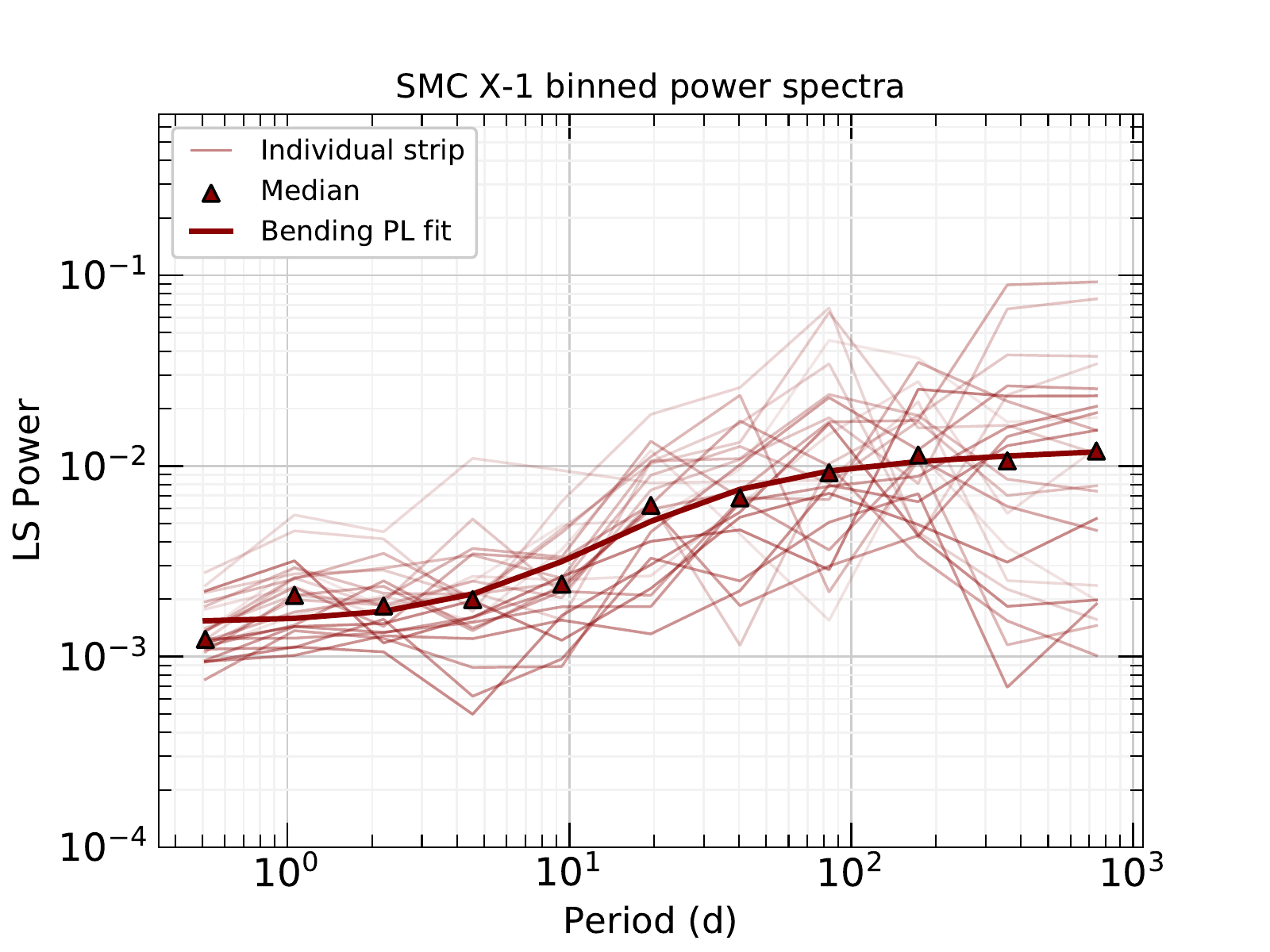}
\includegraphics[width=8cm]{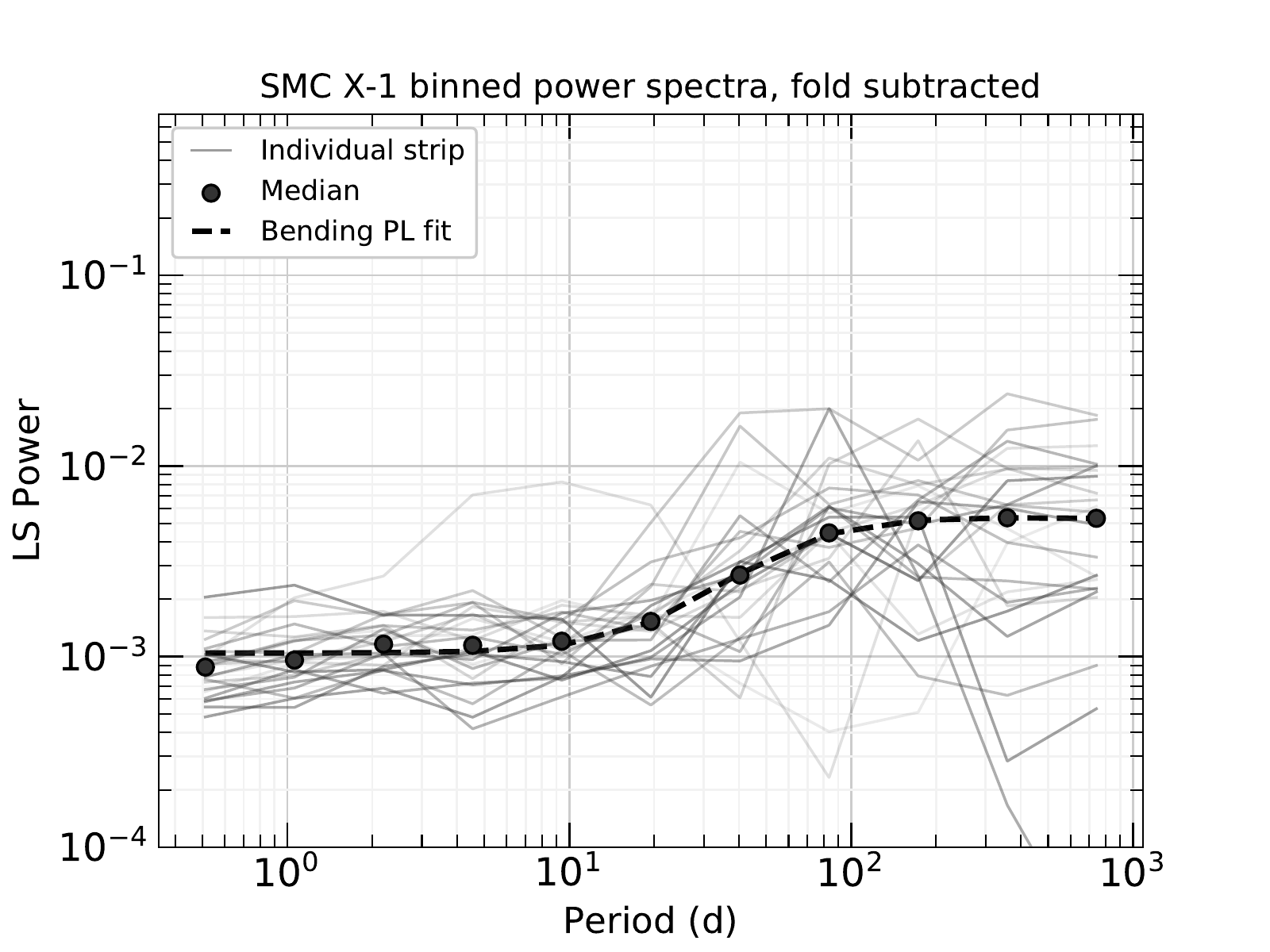}
\caption{Continuum power spectra for the first 24 lightcurve strips (the 25th and final strip shows such sporadic time coverage that its utility for period determination is low). In each panel, the thin lines show the continuua for individual strips, while the symbols and thick line show the median continuum power spectrum over the strips, and a fit to the median continuum power spectrum, respectively. The {\it top panel} shows the binned power spectrum, while the {\it lower panel} shows the result after superorbital and orbital modulation is subtracted. See Section \ref{app:ss:continuum}.}
\label{fig:app:continuumChar}
\end{figure}

 The strips show substantial variability in the indicated continuum (Figure \ref{fig:app:continuumChar}). In some cases, subtraction of the superorbital modulation leaves behind a timescale-dependent "red-noise" continuum with weaker timescale dependence than before. In some others, removal of the superorbital modulation almost removes the timescale dependence of the remaining power spectrum altogether. This might indicate that the remnant noise continuum may be substantially due to intrinsic period and/or profile variation - which would manifest as apparent "red-noise" when a constant-profile, constant-period modulation were subtracted.

To characterise period uncertainty, the median over all the strips of the periodicity-subtracted power spectrum was used. This median continuum power spectrum is reasonably well-fit by a Bending Power-Law (BPL), of the form \citep{connollyDELC}

\begin{equation}
f(\nu) = A \nu^{-a_{\rm lo}} \bigg[
1 + \left( \frac{\nu}{\nu_{\rm bend}}\right)^{a_{\rm hi}-a_{\rm lo}}
\bigg]^{-1} + c
\label{eq:bpl}
\end{equation}

 \noindent where $\nu = 1/P$. Based on the fit to the fold-subtracted continuum power spectrum (Figure \ref{fig:app:continuumChar}, bottom panel), for the model parameters $\{A, a_{\rm lo}, a_{\rm hi}, \nu_{\rm bend}, c\}$ we adopt the values listed in the right-most column of Table \ref{tab:app:bpl} to produce the synthetic datasets. (In practice, while we find $a_{\rm lo} \approx 0$, we retain the full expression as implemented in the {\tt DELCgen.py}~simulation methods of \citealt{connollyDELC}.)

\begin{table}
\begin{center}
\begin{tabular}{ccc}
\hline
Parameter & Raw & Fold-subtracted \\
\hline 
$A$ & $6.84 \times 10^{-3}$ & $5.01 \times 10^{-3}$ \\
$a_{\rm lo}$ & $6.32 \times 10^{-2}$ & $-2.39 \times 10^{-2}$ \\
$a_{\rm hi}$ & $1.56$ & $2.21$ \\
$\nu_{\rm bend}$ & $4.33 \times 10^{-2}$ & $1.94 \times 10^{-2}$ \\
$c$ & $1.52 \times 10^{-3}$ & $1.04 \times 10^{-3}$ \\
\hline
\end{tabular}
\caption{ Bending power-law noise model (Equation \ref{eq:bpl}) fit to the \rxteasm~binned power spectrum. The rightmost column represents the power spectrum after subtraction of the fold to the superorbital and orbital modulations, and was used to generate synthetic datasets for the estimation of superorbital period measurement uncertainty. See Appendix \ref{app:ss:continuum}.}
\label{tab:app:bpl}
\end{center}
\end{table}

\subsection{Comparison of period uncertainties}\label{app:ss:periodUncty}

 Here we present the comparison between the superorbital period uncertainty in the full BPL-noise case and the simpler case assuming timescale-independent white-noise, as well as comparing our simulation-based uncertainties with two standard prescriptions in the literature.

\begin{figure}
\includegraphics[width=8cm]{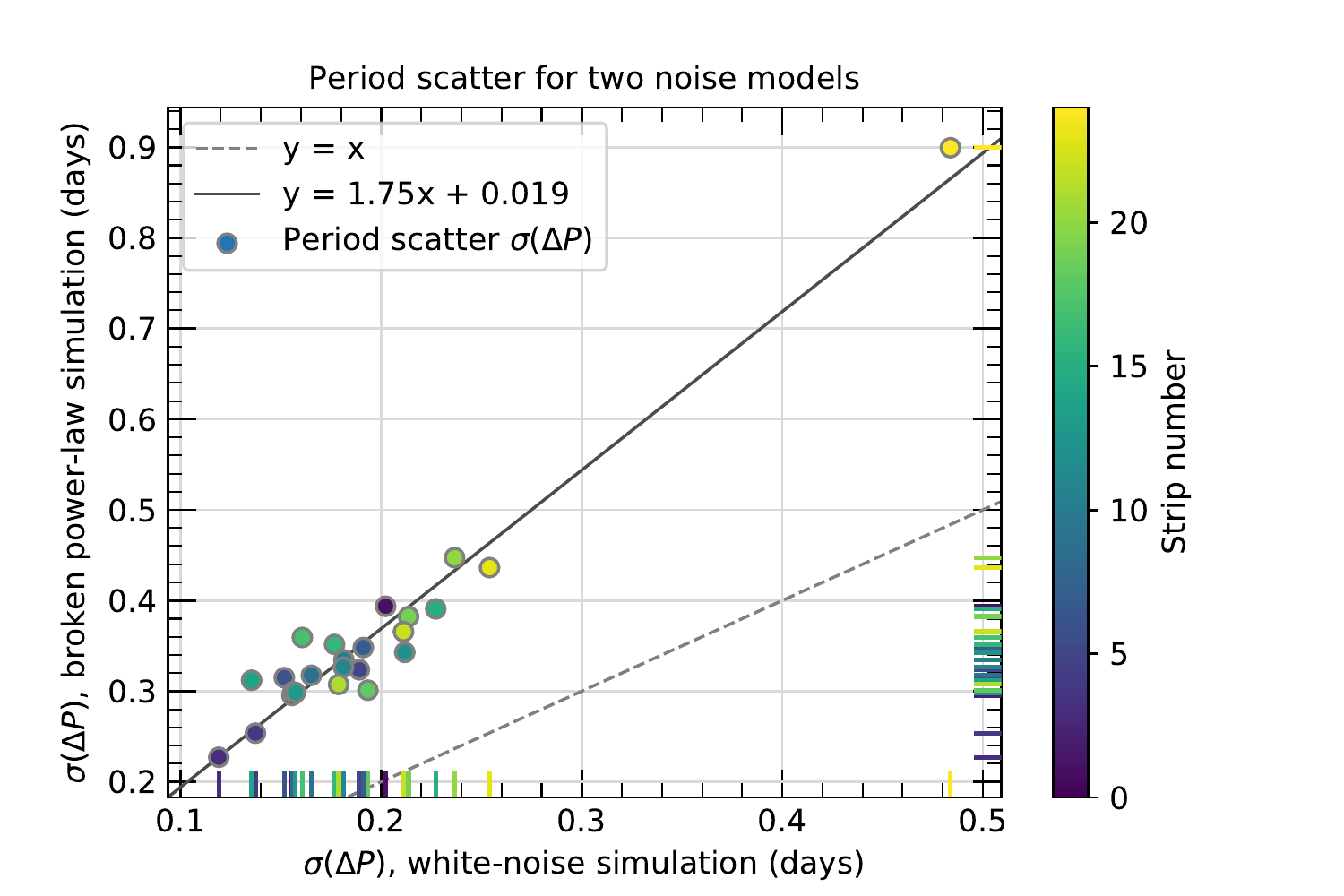}
\includegraphics[width=8cm]{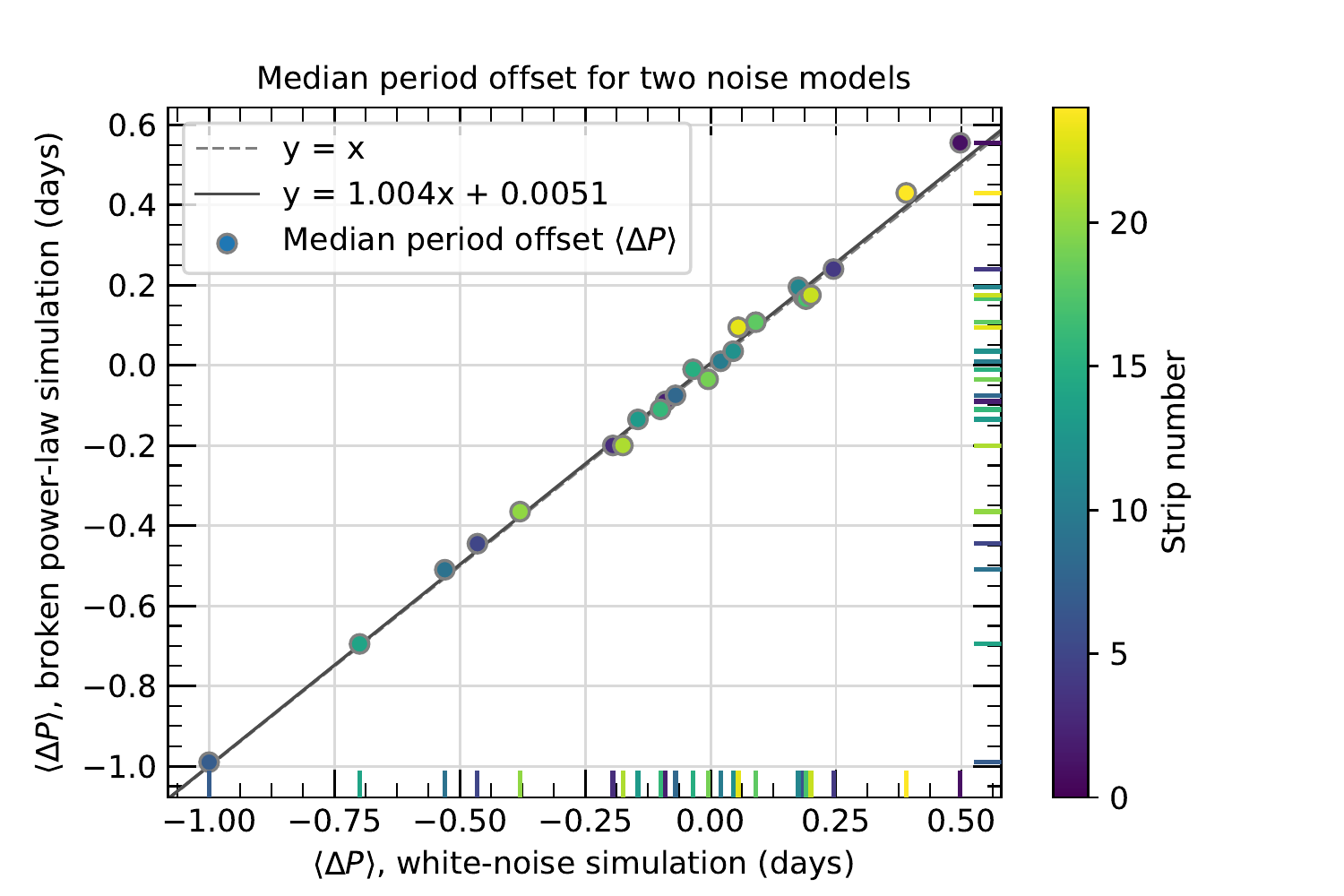}
\caption{ Comparison of the rms ({\it top panel}) and median ({\it bottom panel}) of the offset $\Delta P = $~P(measured)-P(injected), for simulations employing two models of the underlying continuum: a broken power-law (vertical axes) and timescale-independent white-noise (horizontal axes). The plot symbols are color-coded by strip number (strip zero being the earliest). In the top panel, the line corresponding to identical  rms($\Delta P$)~is indicated with a dashed line. See Section \ref{app:ss:periodUncty}.}
\label{fig:app:periodUncty}
\end{figure}

 Figure \ref{fig:app:periodUncty} shows the Monte Carlo estimates of random and median offset $\Delta P$ = P(measured) - P(injected)~thus obtained, for the BPL continuum model and for a white-noise model. For both noise models, the typical random uncertainty in the superorbital period is close to $\sigma(\Delta P) \approx 0.35$~days for the BPL noise model, and close to $\sigma(\Delta P)\approx 0.20$~days for the white-noise model. This is qualitatively as expected, since under a BPL noise model, the continuum surrounding the main power spectrum peak is stronger. While period uncertainties are small in the presence of both types of noise, the impact of the underlying continuum on the detection precision is substantial - detection of a $\sim 50$-day signal over a BPL continuum shows about 1.75 times the period uncertainty as it does over a white-noise continuum.

 Under both noise models, the detected superorbital period is also biased, typically by $\lesssim 0.25$~days but for some strips by as much as $-1.0$~days (see the bottom panel of Figure \ref{fig:app:periodUncty}; note that this is {\it not} the same bias as the Gaussian-peak bias presented in Appendix \ref{app:ss:peakBias}). This bias is indistinguishable between the BPL and white-noise cases and in most cases is substantially larger than the random uncertainty. The bias probably represents the impact of the window function due to the uneven sampling \citep[e.g.][]{scargle82} and is reproducible, allowing us to correct for it when estimating the superorbital period for each lightcurve strip (Section \ref{sssec:rednoise}).

We may also compare the $\sim 0.3$-day estimate for the superorbital period uncertainty with commonly-used formal estimates (e.g. \citealt{horneBaliunas1986}). The conservative upper-limit for frequency uncertainty from a signal that is measured over total time interval $T$, i.e.
\begin{equation}
\delta f_2 = \frac{1}{2T} 
\label{eq:unctyFormal:conservative}
\end{equation}
translates to period uncertainty $\sigma(P) \approx 7$~days, much larger than our simulation-based estimate, as expected.

 For completeness, we also consider the widely-used period uncertainty estimate of \citet[][ itself derived from \citealt{kovacs81}]{horneBaliunas1986}, which predicts the frequency uncertainty under a model of a single sinusoidal signal over Gaussian white noise. As presented in \citet{levine2011}, this prescription takes the form
\begin{equation}
\delta f_1 = \frac{3}{8}\frac{1}{T\sqrt{\mathcal{P}_r}}
= \frac{3\delta f_2}{4}\sqrt{\frac{\langle \mathcal{P} \rangle}{\mathcal{P}_{\rm raw}}} \equiv K\delta f_2
\label{eq:unctyFormal:levine}
\end{equation}
 In the first form of $\delta f_1$, $\mathcal{P}_r$~is the peak power in the power spectrum, and the power spectrum has first been normalized so that the average power is unity \citep{levine2011}. In the second, $\mathcal{P}_{\rm raw}$~is the peak period in the power spectrum before normalization, and $\langle \mathcal{P} \rangle$~is the appropriate average power in the periodogram.

 Expression (\ref{eq:unctyFormal:levine}) was originally derived for the white-noise case \citep{kovacs81}, and the presence of timescale-dependent red-noise complicates the determination of the normalizing factor $\langle \mathcal{P} \rangle$. For a given strip, the equivalent normalization is likely in the range $10^{-3} \lesssim \langle \mathcal{P} \rangle \lesssim 2 \times 10^{-2}$~d (Figure \ref{fig:app:continuumChar}). With raw periodogram peaks in the range $0.1 \lesssim \mathcal{P}_{\rm raw} \lesssim 0.35$~d, this suggests ($0.04 \lesssim K \lesssim 0.33$), or, finally, ($0.3 \lesssim \delta f_1 \lesssim 2.4$)~d. While the lower bound of this range is consistent with the full simulations, it corresponds to normalization by the short-timescale end of the underlying noise-continuum. Strict application of the estimate (\ref{eq:unctyFormal:levine}) to the \rxteasm~power spectrum of SMC X-1 is not well-justified and we caution against its use in red-noise cases. We prefer to quote uncertainties directly from the synthetic lightcurves that we have tailored to match the observed behaviour of SMC X-1.

\subsection{Dependency of measurement uncertainty on strip-length $\tau$}\label{app:ss:vsTau}

\begin{figure}
\includegraphics[width=8cm]{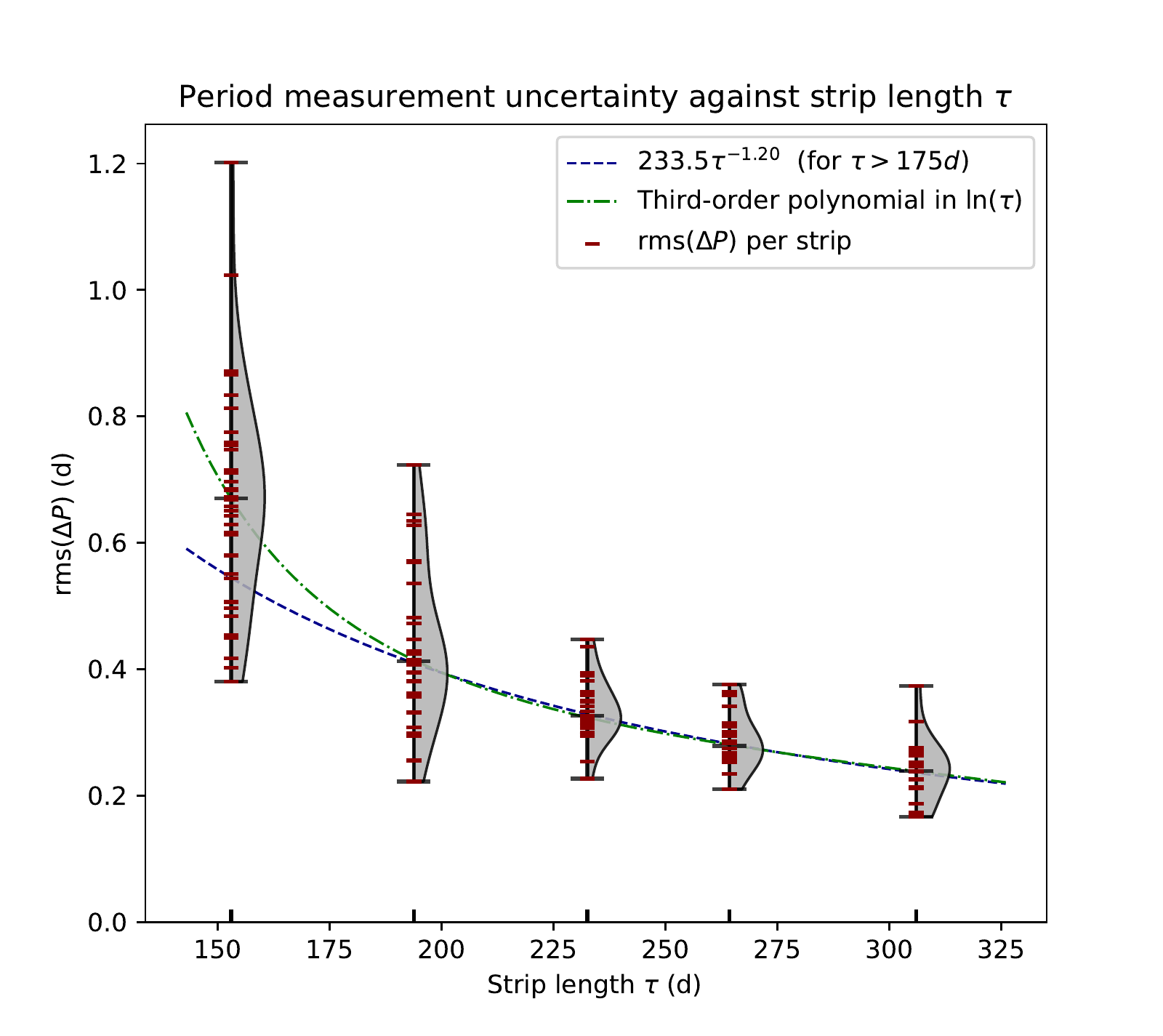}
\caption{ Evolution of the superorbital period measurement uncertainty as the strip-length $\tau$~is varied. Dark red points show the rms of $\Delta P$~(i.e., measured minus simulated period) for each strip with end date prior to MJD 55,600 (at which point \rxteasm~coverage becomes sparse; Figure \ref{fig:lcurve}). The gray shaded areas visualise the distribution of rms($\Delta P$) for each value of $\tau$, with the median and extrema indicated with black horizontal lines in each case. The black vertical notches on the horizontal axis indicate the strip-length $\tau$~in each case. The blue dashed line indicates a power-law fit to the median rms$(\Delta P)$~for $\tau > 175$~d, while the green dot-dashed line indicates a polynomial fit to $\ln(\tau)$~for all strip-lengths in the Figure. See Appendix \ref{app:ss:vsTau}.}
\label{fig:app:vsTau}
\end{figure}

 Figure \ref{fig:app:vsTau} presents the variation of superorbital period measurement uncertainty as the strip-length $\tau$~is varied. For each strip, all the methods of Appendices \ref{app:ss:peakBias} and \ref{app:ss:continuum} were applied and the uncertainty in $\Delta P$~characterised in the same manner. There is a weak dependence on the strip-length, with the longest strip-length affording the smallest measurement uncertainty, on the assumption that the superorbital period is constant throughout the strip. For $(175 \lesssim \tau \lesssim 325)$~days, the median value of the measurement uncertainty for each value of $\tau$~seems to be well-represented by a power-law of the form $\langle$rms($\Delta P$)$\rangle \approx 234\tau^{-1,2}$, while shorter strips suffer from larger uncertainty than predicted by this trend. The strip-length $\tau = 230$~d adopted in the main paper (i.e., breaking the \rxteasm~dataset into 25 equal strips) presents a good compromise between minimising measurement uncertainty and retaining sensitivity to variation in superorbital period.

\section{Increasing the speed of red-noise lightcurve simulation}
\label{ap:timing}

When conducting red-noise Monte Carlo tests (Section \ref{sssec:rednoise}) with the {\tt DELightcurve} suite, we found large reproducible variation (by up to a factor 90) between the strips in the amount of time required to generate and analyse the synthetic lightcurves. We show here that this is most likely due to the execution time for the inverse Fast Fourier Transform (iFFT), and its dependence on the factorization properties of the length of a particular intermediate array in the \citet{connollyDELC} implementation of the \citet{timmerKoenig1995} method. Since a single time interval $\tau$~per strip is unlikely to allow convenient array-lengths for all strips when the data are unevenly sampled, we present a simple workaround to remove the strong non-uniformity in execution time, which removes a bottleneck to practical generation of sufficient numbers of synthetic lightcurves for Monte Carlo analysis. 

In tests, we traced the bottleneck to the {\tt TimmerKoenig} method within the {\tt DELCgen.py} module \citep{connollyDELC}.\footnote{f Note that this bottleneck is likely a feature of the algorithm itself and not the {\tt DELCgen.py} implementation.} A step within this method takes the iFFT of an array of length $L \sim 100$~times the input data length (extending the lightcurve length by a large factor to account for spectral leakage). Since in general the length of this large-array will not be a power of two, we might expect the time elapsed in taking the iFFT to scale with the square of the number of elements in the longest partition of the length-$L$~array (usually the largest prime factor of $L$, e.g. \citealt{press02}).

\begin{figure}
\includegraphics[width=10cm]{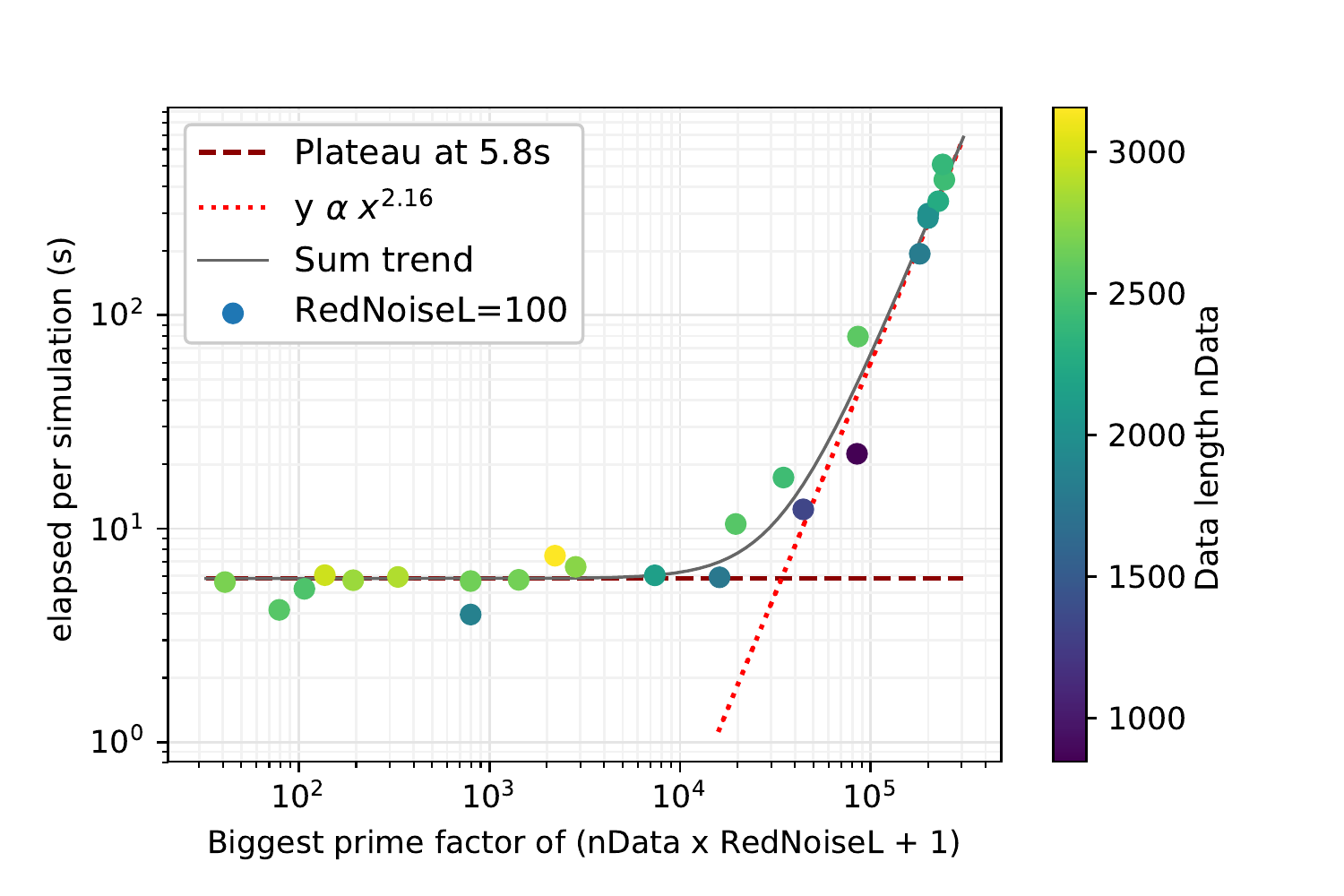}
\caption{Average elapsed time per synthetic lightcurve (generation plus measurement and recovery of superorbital period) with periodicities over red-noise (Section \ref{sssec:rednoise}), for each of 25 equal-length strips using the RXTE/ASM sample times of SMC X-1 and the parameter {\tt RedNoiseL} fixed at its default value of 100. Durations are plotted as a function of the largest prime factor $x$~of the length $L$~of the finely-sampled array appropriate for each strip, and the color scale shows the number $N$~of datapoints in each strip. For $x \gtrsim 10^4$~elements, the elapsed time scales approximately as $\propto x^2$~(we find $\propto x^{2.16}$). These timings were measured on a 1.8 GHz laptop with 8GB RAM, using {\tt DELCgen.py} as last modified 2016 December 01 (the latest version as of 2017 December). See Appendix \ref{ap:timing} for discussion.}
\label{fig:durationStatic}
\end{figure}

Within the module {\tt DELCgen.py}, the length $L$~of the finely-sampled array is constructed from the number of input datapoints $N$~and a length-parameter {\tt RedNoiseL}, via the relation $L = (N \times {\tt RedNoiseL}) + 1$. As shown in Figure \ref{fig:durationStatic}, for a uniform value of {\tt RedNoiseL} for all strips, the execution time is dominated at the long end by an approximate $\propto x^2$~scaling, where $x$~is the largest prime factor of the array-length $L$. This supports the hypothesis that it is the partitioning of data for the iFFT computation that dominates the total execution time.

Since the parameter {\tt RedNoiseL} is user-controlled (with default value 100), it allows the large-array length $L$~to be tuned for performance. We follow a simple heuristic, looping through trial values in the range $90 \le {\tt RedNoiseL} \le 105$~and picking the value that minimizes the largest prime factor $x$, using the Python package {\tt sympy}\footnote{\url{http://www.sympy.org/en/index.html}} to perform the factorization. 

\begin{figure}
\includegraphics[width=10cm]{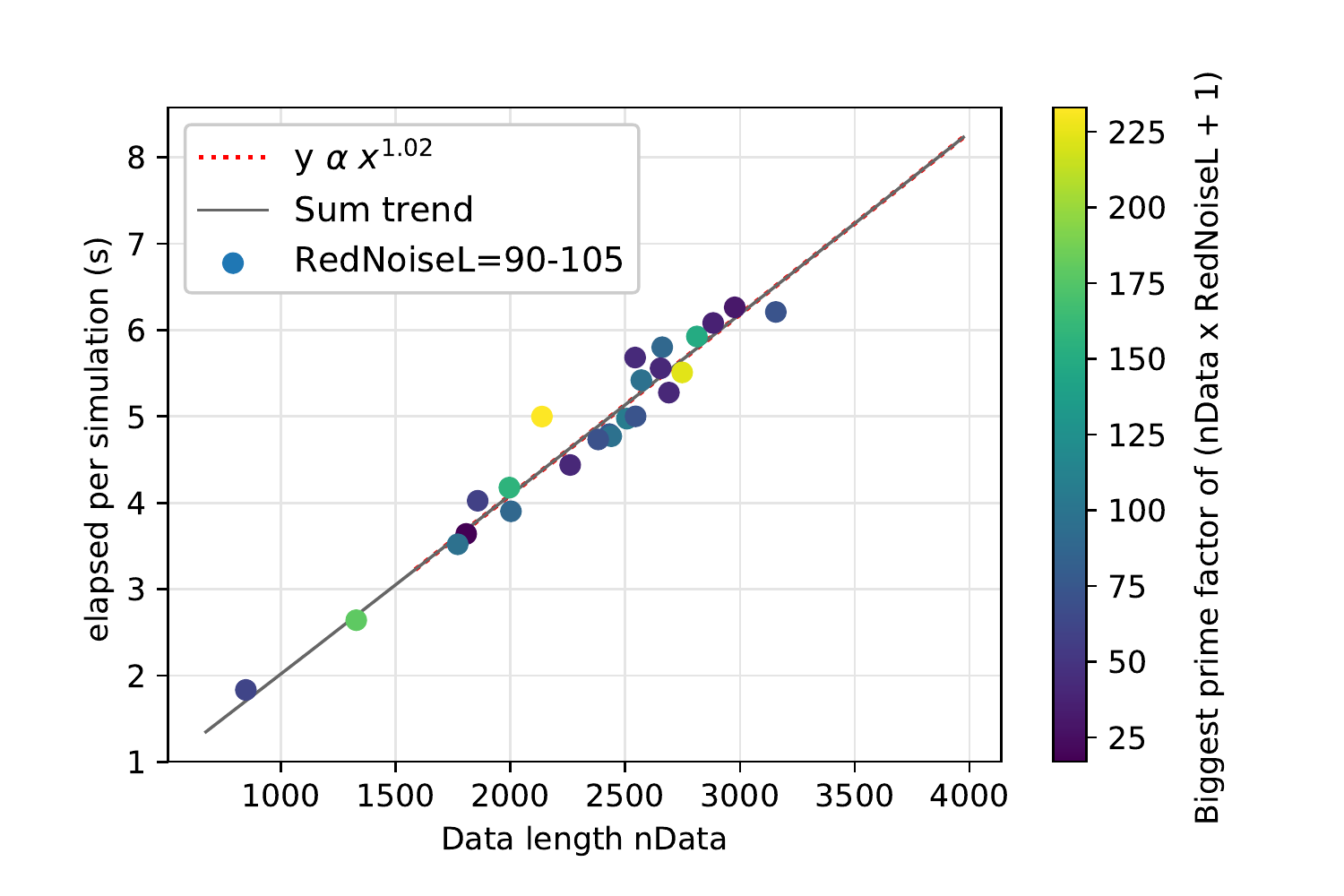}
\caption{Average elapsed time to simulate and analyse a single synthetic lightcurve (similarly to Figure \ref{fig:durationStatic}), but plotted against the number $N$~of datapoints in each strip and with the color scale showing the largest prime factor of $L$~for each strip. This run differs from that presented in Figure \ref{fig:durationStatic} in that for each strip the oversampled array-length $L$~was dynamically tuned (via the parameter {\tt RedNoiseL}) to minimize the largest prime factor of $L$~and thus the elapsed time. See Appendix \ref{ap:timing} for discussion.}
\label{fig:durationDynamic}
\end{figure}

Figure \ref{fig:durationDynamic} shows the improvement in execution time and its uniformity when {\tt RedNoiseL} is dynamically varied rather than set equally for all simulations. No $\propto x^2$~dependence was found, and instead the duration now depends most strongly on $N$, the number of points in the original dataset that lie within the strip. For the most problematic strips, the execution time was reduced from $\sim 500$~seconds to $\sim 5-6$~seconds, now likely dominated by the measurement process applied to each simulation and scaling with $N$.

In our tests, this parameter-tuning procedure reduces the total time required to conduct our Monte Carlo simulations by about a factor 20, at a cost of slight non-uniformity in the length-multiplication factor {\tt RedNoiseL}~between the strips. If it is particularly important that {\tt RedNoiseL} be as uniform as possible over the strips, then more sophisticated approaches might be considered, including walking out from the default value of 100 until $x$~drops below an acceptable threshold determined from a prior run of the simulations. For the present, however, we consider the simple heuristic used here to be an acceptable compromise between uniformity and execution time.

\bsp	
\label{lastpage}
\end{document}